# Multifaceted neural representation of words in naturalistic language

Xuan Yang [a,b], Chuanji Gao [c], Cheng Xiao [d], Nicholas Riccardi[b,e], Rutvik H. Desai [a,b,*]

[a] *Department of Psychology, University of South Carolina, Columbia, SC, USA*

[b] *Institute for Mind and Brain, University of South Carolina, SC, USA*

[c] *Department of Psychology, Nanjing Normal University, Nanjing, Jiangsu, China*

[d] *Linguistics Program, University of South Carolina, Columbia, SC, USA*

[e] *Department of Communication Science and Disorders, University of South Carolina, Columbia, SC, USA*

Xuan Yang: xy6@email.sc.edu

Chuanji Gao: chuanji.gao@outlook.com

Cheng Xiao: cxiao@email.sc.edu

Nicholas Riccardi: riccardn@email.sc.edu

Rutvik H. Desai: rutvik@sc.edu

Correspondence should be addressed to Rutvik H. Desai

## Abstract

Understanding how the brain represents the multifaceted properties of words in context is essential for explaining the neural architecture of human language. Here, we combine large-scale psycholinguistic modeling with naturalistic fMRI to uncover the latent structure of word properties and their neural representations during narrative comprehension. By analyzing 106 psycholinguistic variables across 13,850 English words, we identified eight interpretable latent dimensions spanning lexical usage, word form, phonology–orthography mapping, sublexical regularity, and semantic organization. These factors robustly predicted behavioral performance across lexical decision, naming, recognition, and semantic judgment tasks, demonstrating their cognitive relevance. Parcel-based and multivariate fMRI analyses of narrative listening revealed that these latent dimensions are encoded in overlapping yet functionally differentiated cortical systems. Multidimensional scaling and hierarchical clustering analyses further identified four interacting subsystems supporting sensorimotor grounding, controlled semantic retrieval, resolution of lexical competition, and contextual–episodic integration. Together, these findings provide a unified neurocognitive framework linking fundamental lexical psycholinguistic dimensions to distributed cortical systems engaged during naturalistic language comprehension.

# Introduction

Understanding how the brain represents word meaning and form is fundamental to explaining the neural bases of language. Properties of words can be broadly classified into semantic (related to meaning), orthographic (related to the written form), and phonological (related to the spoken or signed form). Additionally, more general properties such as frequency apply to all three categories, and some properties such as consistency relate to the interaction of these categories. These properties have been robustly identified across languages, with each contributing uniquely to word recognition [1–6]. However, the ways in which they are selectively represented and jointly organized in the brain remain poorly understood. Disentangling their overlapping structure and characterizing how they are distinctly and interactively encoded in the brain are essential for advancing theoretical models of word processing and informing targeted interventions for language-related disorders.

A substantial body of research has examined the neural representation of psycholinguistic properties using single-word stimuli in highly controlled paradigms [7–15]. A strength of these approaches is the ability to examine specific lexical variables by removing complex effects of semantic and syntactic context, and factoring out correlated variables, albeit in a typically small number of stimuli and in specific task contexts. A complementary approach is naturalistic listening or reading tasks that allow participants to engage with language in a manner more closely aligned with real-world communication [16–18]. This ecologically grounded approach has been shown to recruit a broader, bilateral network during narrative comprehension [19–24], underscoring the value of studying lexical-level neural representations in naturalistic settings that better approximate everyday language use.

A fundamental issue facing both single-word and naturalistic approaches is that words inherently possess numerous intercorrelated properties, making it difficult to disentangle the unique effect of individual psycholinguistic features [25]. This problem has been brought into greater focus with the proliferation of large-scale megastudies providing norm- and corpus-based ratings for a vast array of lexical features [1,4,26–36]. Whereas earlier research was constrained by the limited availability of such measures, contemporary studies now face the opposite challenge, a surfeit of candidate variables that complicates interpretation and replication. As the number of variables increases, it becomes increasingly difficult to design stimulus sets that adequately decorrelate semantic, phonological, and orthographic properties [11].

In addition, nominally similar constructs are often operationalized in multiple ways. For example, English word frequency has been estimated using over 15 different corpora [27,37–40], yielding subtly different values and contributing to inconsistent findings across behavioral and neuroimaging studies [27]. Variables sharing the same label may also capture distinct underlying constructs. For instance, semantic neighborhood density (SND) computed from vector-based semantic similarity reflects both thematic and taxonomic relationships, whereas SND derived from word association norms primarily reflects thematic relationships [34,41].

Taken together, a more comprehensive characterization of the latent structure underlying interrelated psycholinguistic properties in naturalistic language contexts can help disentangle conceptual ambiguities and advance our understanding of their neural correlates. In the present study, we analyze an extensive set of lexicosemantic variables across the largest number of words aggregated from major psycholinguistic databases, aiming to identify latent dimensions that structure psycholinguistic space. We then examine how these dimensions are represented in the brain during naturalistic word processing, using fMRI data collected while participants listened to spoken narratives. This study provides a novel framework for linking latent psycholinguistic structures to brain responses under conditions that closely approximate real-world language comprehension.

## Results

To investigate how psycholinguistic properties are organized and represented in the brain during naturalistic language comprehension, we conducted a multi-stage analysis combining large-scale lexical data, behavioral norms, and brain activity during narrative listening. We first applied exploratory factor analysis (EFA) to an extensive set of psycholinguistic variables to identify latent factors underlying word properties. We then assessed the cognitive relevance of these factors using behavioral outcome measures from single-word processing tasks [42]. Finally, we examined how those factors are uniquely encoded and integrated in the brain using both parcel-based parametric modulation analyses and data-driven approaches on fMRI data collected during narrative listening.

**Eight latent factors capture the structure of lexical form and meaning**

Given the vast number of highly intercorrelated psycholinguistic variables, we used EFA to uncover the fundamental latent factors. This approach allows us to characterize the latent structure of psycholinguistic properties while accounting for correlations among factors in higher-dimensional space. we selected 106 variables from 20 databases (see Table S1) from the South Carolina psycholinguistic metabase (SCOPE) [42], one of the most comprehensive and curated collection of psycholinguistic properties, spanning General, Phonological, Orthographic, Semantic, and Morphological categories. While SCOPE contains more than 200 variables, this number was chosen to balance the competing goals of maximizing the number of variables and maximizing the number of words for which all variable values are available. The final dataset consisted of 13,850 unique words (5,905 unique lemmas). Correlations between variables (Fig. 1) shows clustering, suggesting potential latent components. Based on the scree plot (Fig. 2A), we selected an eight-factor model that explained more than half (56.83%) of the total variance [43].

Based on factor loadings across the 106 variables (Fig. 2C and Table S2), the eight extracted factors were interpreted as follows (see Table S3 for example words and lemmas with the highest factor scores). FA1 (*Frequency*) was associated with frequency and contextual diversity norms from multiple corpora, as well as age of acquisition and semantic diversity, reflecting amount of usage in diverse contexts. FA2 (*OPNeighborhood/Length*) was associated with the number of letters, phonemes, and syllables, as well as with orthographic and phonological neighborhood density, reflecting the length and neighborhood density of both phonological and orthographic forms. FA3 (Phonology-to-Orthography Consistency, *POconsistency*, also known as feedback, sound-to-spelling, or orthographic consistency) and FA4 (Orthography-to-Phonology Consistency, *OPconsistency,* also known as feedforward, spelling-to-sound, or phonological consistency) captured the regularity and consistency of bidirectional mapping between graphemes and phonemes. FA5 (Graphotactic/phonotactic probability, *GPprobability*) loaded on graphotactic and phonotactic probabilities, indicating how frequent a word's letter or sound sequences are. FA6 (*Concreteness*) was strongly linked to concreteness, sensory-relatedness, and imageability, indicating whether a concept can be experienced with the senses and the ease with which a mental image can be formed. FA7 (*TaxonomicSND*) [1] reflected taxonomic semantic neighborhood density derived from co-occurrence-based metrics and hierarchical similarity. Finally, FA8 (*AssociativeSND*) represented associative neighborhood density based on word association norms, capturing contextual and experiential associations.

**Latent psycholinguistic factors predict behavioral performance across multiple tasks**

We next examined whether these eight latent factors reflect meaningful behavioral relationships in word processing rather than arbitrary statistical patterns. Specifically, we assessed their correlations with reaction times (RT) and accuracy across five psycholinguistic tasks aggregated from seven databases [42], including auditory and visual lexical decision, word naming, semantic decision, and word recognition (Fig 3 and Table S1). All eight factors exhibited significant zero-order correlations with task performance (Fig. 3A), and many of these correlations remained robust after controlling for the variance shared with other factors (Fig. 3B).

---

[1] For easier comparison, we recoded the TaxonomicSND (FA7) factor scores by multiplying the original factor scores by −1, such that higher values on the factor scores of both SND factors reflect denser semantic neighborhood.

Higher **FA1 (Frequency)** scores (high frequency) were consistently correlated with faster RTs and higher accuracy in visual lexical decision, word naming, and auditory lexical decision. In semantic decision, this effect was not significant after controlling for other factors (Fig. 3B). There was a reversed effect in word recognition task, with lower scores (low frequency) correlated with better recognition performance. **FA2 (OPNeighborhood/Length)** was related to slower reaction times and higher accuracy, particularly in visual lexical decision and word naming, suggesting that words that are longer and have sparser orthographic/phonological neighborhood are processed longer but more accurately. **FA3 (POconsistency)** showed weaker and task-specific associations. Words that have inconsistent phonology-to-orthography mappings had a facilitating effect on processing speed in lexical decision, a similar effect on both processing speed and accuracy in semantic decision, but an interference effect on processing speed in word naming. **FA4 (OPconsistency)** had the opposite effect: Words that have inconsistent orthography-to-phonology mappings had an interference effect on both processing speed and accuracy in lexical decision and word naming, but not semantic decision. **FA5 (GPprobability)** was positively correlated with reaction times and negatively correlated with accuracy across tasks, indicating that words that have less typical or less predictable letter and sound patterns were processed more slowly and less accurately. **FA6 (Concreteness)** correlated negatively with reaction times and positively with accuracy across tasks, consistent with previous findings on concrete and imageable word advantages. **FA7 (TaxonomicSND)** and **FA8 (AssociativeSND)** showed consistent but opposite effects on word processing. Words with denser taxonomic SND had an interfering effect, while words with denser associative SND had a facilitating effect across lexical decision, semantic decision and word naming tasks.

To further assess the predictive utility of the latent factors, we compared their ability to predict 15 behavioral outcome variables across five different tasks against that of the "best single predictor" among individual psycholinguistic variables (see Methods for details). This analysis tests whether the predictive strength of the latent factors is comparable to that of traditional isolated variables commonly used in neuroimaging research, ensuring that our behavioral and neural findings can be meaningfully related to earlier work. All eight factors either outperformed or showed no significant difference from the "best single predictor" in predicting behavioral outcomes for at least two behavioral variables (Table S3), demonstrating that the latent factors provide a reliable and comparable basis for examining neural representations in the following analyses.

**Latent psycholinguistic factors engage distinct and overlapping cortical systems during naturalistic listening**

We next addressed the central question of how human brain represents those latent dimensions of word properties during naturalistic language comprehension. We conducted a parcel-based parametric modulation analysis using data from the Narratives collection [44], comprising 102 young adults listening to 10 unique narratives (213 fMRI scans). Because our goal was to characterize factor-related representations at the level of distributed cortical organization rather than fine-grained local activations, we analyzed responses within anatomically and functionally defined cortical parcels. This approach aggregates voxel-wise responses within each parcel, improving statistical reliability and facilitating interpretation in relation to established large-scale brain networks, while enabling direct comparison of shared and distinct representations across the eight latent psycholinguistic factors. Specifically, in the first-level analysis, all eight latent factors were entered simultaneously into a single general linear model (GLM) for each voxel, enabling us to estimate the unique neural contribution of each factor while controlling for the others. For each factor and each fMRI scan, we averaged statistical values across the voxels within each of the 360 parcels (180 per hemisphere) of the HCP-MMP atlas [45] (see Table 1 for the descriptions of sample parcels). At the second level, one-sample t-tests were performed for each factor to generate whole-brain statistical maps reflecting group-level parcel-wise modulation by the corresponding factor. Significance was examined at $p < .005$, with FDR correction applied for multiple comparisons. As a validation, we also conducted voxel-wise analysis which yielded activation patterns that spatially converged with parcel-based analyses (Fig S1).

FA1 (Frequency) was negatively correlated with activation in bilateral inferior frontal sulcus (IFS) and superior temporal gyrus (STG), as well as left inferior frontal junction (IFJ), middle temporal gyrus (MTG), inferior temporal gyrus (ITG), and fusiform gyrus (FG). FA1 also positively correlated with activation in bilateral insula and parieto-occipital sulcus (POS), as well as right posterior STG, anterior angular gyrus (AG), posterior cingulate cortex (PCC), and precuneus.

FA2 (OPNeighborhood/Length) showed widespread negative correlations, including bilateral superior and middle frontal gyrus (SFG, MFG), dorsal medial prefrontal cortex (DMPFC), inferior frontal gyrus (IFG), STG, STS, MTG, supramarginal gyrus (SMG), and AG. Positive correlations were observed in bilateral PCC and precuneus.

FA3 (POconsistency) was negatively correlated with activation in bilateral DMFPC, IFG, STG, STS, and SMG. Positive correlations were observed in left POS.

FA4 (OPconsistency) was negatively correlated with activation in left ITG, fusiform, and precuneus, and positively correlated with activation in bilateral IFG, left posterior insula, as well as right IFS, STG, and STS.

FA5 (GPprobability) was negatively correlated with activation in bilateral precuneus and PCC and positively correlated with activation in right anterior insula. Complementary voxel-wise analyses revealed additional positive correlations in bilateral IFG Pars Triangularis (IFG-Tri), inferior parietal lobule (IPL), and SFG.

FA6 (Concreteness) exhibited an extensive and distributed pattern, with negative correlation in bilateral IFG, STG, STS, MTG, and lateral occipital cortex (LOC), as well as left posterior MFG (parcel 55b) and PCC. Positive correlations were observed in bilateral IFS, IFJ, SMG, posterior MTG, temporo-parieto-occipital junction (TOPJ) fusiform, ITG, parahippocampal gyrus (PHG), POS, PCC, dorsal anterior cingulate cortex (dACC), as well as left primary motor cortex and somatomotor cortex.

FA7 (TaxonomicSND) was negatively associated with activation in left LOC. Positive correlation was observed in bilateral IFS, IFJ, left posterior STG/SMG, ITG, right MFG, STG, STS, and precuneus.

FA8 (AssociativeSND) was negatively correlated with activation in bilateral posterior MTG/TOPJ, PCC, and left PHG, and positively correlated with activation in bilateral ventral medial prefrontal cortex (VMPFC), PCC, and POS. Complementary voxel-wise analyses revealed additional positive correlations in bilateral IFG-Tri and STG.

**Shared and specialized cortical systems jointly support multidimensional processing of words**

The parcel-wise GLM analyses revealed distinct activation patterns for each latent dimension while controlling for the others. However, substantial overlap was also observed, with regions such as the IFG, STS, and precuneus being modulated by multiple factors. To more systematically distinguish between integrative and specialized cortical responses, we performed a parcel-based conjunction analysis to identify potential hub-like regions engaged across several aspects of word processing, in contrast to regions selectively tuned to a single psycholinguistic factor. We use the term representational profile to describe a parcel's pattern of modulation across the eight latent psycholinguistic factors.

Several regions were broadly modulated by multiple latent factors showing a similar representational profile (Fig 5B), including bilateral posterior medial SFG (parcel SFL), IFG and IFS around Pars Triangularis (parcel 45 and IFSa), POS (parcel POS2), left posterior MTG (parcel PHT), ITG (parcel TE2p), right STG/STS (parcel A5, STSdp, and STSvp) extending to SMG and AG (parcel STV and PSL), PCC (parcel RSC) and precuneus (parcel 7m). The right STG had the most overlap (seven factors), followed by right STS (six factors), and then by left posterior MTG and ITG as well as right IFG/IFS and precuneus (four factors).

The regions selectively modulated by one or two factors were more distributed and exhibited distinct representational profiles (Fig 5A and Fig S2). Some regions were mainly modulated by factors associated with semantic-related factors (FA6 Concreteness, FA7 TaxonomicSND, and FA8 AssociativeSND), including VMPFC,

IFS/IFJ, TPOJ, posterior to middle cingulate cortex (MCC; BA 23), dorsal anterior cingulate cortex (dACC; BA24), PHG, and anterior FG (parcel TF), whereas some regions showed selective sensitivity for factors associated with lexical processing and phonology-orthography mapping (FA1 Frequency and FA4 OPConsistency), particularly in bilateral posterior FG and ventral occipitotemporal cortex (vOTC; parcel FFC and PH).

## Multivariate analyses reveal four functional subsystems underlying word processing during narrative listening

Univariate analyses revealed that some parcels exhibited highly similar representational profiles across the eight latent factors. For example, right A5, STSdp, and left BA45 showed convergent sensitivity to words that are shorter, have denser form-based neighborhoods, exhibit greater phonology-to-orthography inconsistency, are more abstract, and have denser taxonomic and associative semantic neighborhoods. This convergence suggests that these regions may operate as components of a common functional network during word processing.

At the same time, there were clear dissociations between spatially adjacent regions. For instance, left IFG was preferentially sensitive to abstract semantic properties, whereas the adjacent IFS showed greater sensitivity to concreteness. Similarly, although FG/ITG was broadly modulated by frequency, its anterior portion (parcel TF) was sensitive to concreteness, whereas its posterior portion (parcels FFC and TE2p) was selectively sensitive to orthography-to-phonology mapping. These findings indicate that functionally distinct representational profiles can emerge at a fine spatial scale, even within contiguous cortical territory.

While thresholded univariate maps are well suited for identifying regions significantly modulated by individual factors, they are less informative about the similarity structure of representational profiles across regions. Because our univariate analyses employed stringent correction for multiple comparisons, parcels with broadly distributed but moderate effects across factors may show limited thresholded activation despite exhibiting representational profiles similar to other regions. For example, parcel left 55b in the posterior MFG showed a representational pattern closely resembling that of left IFG and right STS across the full factor set, despite reaching significance for only a subset of factors in the univariate maps. To directly characterize functional relationships among cortical parcels based on their multivariate representational profiles, we constructed an eight-dimensional factor-sensitivity vector for each parcel and computed pairwise Euclidean distances between parcels, yielding a parcel-by-parcel distance matrix that was subsequently analyzed using multidimensional scaling (MDS). This approach projects high-dimensional representational structure into a lower-dimensional space, enabling functional distances between parcels to be quantified. Four dimensions emerged, accounting for 86% of the total variance.

Along the first dimension (Fig. 6), bilateral STG, STS, IFG, and DMPFC were functionally distinct from bilateral IFS, IFJ, MFG, TPOJ, PHG, MCC/dACC, and POS. Parcels on the former end of this dimension preferentially encoded words that are more abstract and embedded in denser semantic neighborhoods, whereas parcels on the latter end were more sensitive to words that are concrete, manipulable, and grounded in sensory experience. The second dimension (Fig. 6) separated bilateral PCC and precuneus from bilateral IFS, IFJ, SMG, and anterior FG/ITG. The former group encoded words that are more frequently used, longer, have sparser phonological and orthographic neighborhoods, and denser associative semantic neighborhoods, whereas the latter encoded words that are less frequent, more concrete, and embedded in denser taxonomic semantic neighborhoods.

A third dimension (Fig. 6) distinguished bilateral PCC, precuneus, and right STG from left posterior MTG and ventral occipitotemporal cortex. Parcels in the former group were more sensitive to contextual and associative properties, whereas parcels in the latter group preferentially encoded words with higher phonotactic and graphotactic probability, lower frequency, and greater inconsistency in orthography–phonology mapping. Finally, a fourth dimension (Fig. 6) separated left IFS, IFJ, STG, and middle-to-anterior FG/ITG from right posterior STG and SMG/AG. The former group encoded words that are less frequent but embedded in denser taxonomic and associative semantic neighborhoods, whereas the latter encoded words that are more concrete, more frequent, and shorter.

Together, these results reveal a structured functional organization of cortical parcels with respect to the eight latent psycholinguistic factors, suggesting the presence of distinct functional subsystems. However, MDS provides a continuous embedding of representational similarity and does not explicitly delineate discrete subsystem boundaries or assign parcels to groups. To directly characterize subsystem structure, we applied hierarchical clustering as a complementary, data-driven approach to group parcels based on their multivariate representational profiles. This analysis identified four interpretable subsystems (Fig. 7).

The first subsystem comprised bilateral SMG, TPOJ, PHG, MCC/dACC, POS, and left primary motor and somatosensory cortex, and selectively encoded concrete words with strong sensorimotor and experiential associations. The second subsystem included bilateral IFS, IFJ, middle-to-anterior FG/ITG, left posterior MTG, and superior parietal lobule, and was preferentially engaged by words with properties commonly associated with controlled semantic retrieval, such as low-frequency words, concrete words with rich feature structure, or words embedded in dense taxonomic semantic neighborhoods. The third subsystem encompassed bilateral DMPFC, MFG, IFG, STG, and STS, and encoded words with features that impose greater demands on resolving lexical competition, including those with dense phonological or orthographic neighborhoods or inconsistent phonology–orthography mappings. The fourth subsystem consisted of bilateral PCC, precuneus, and lateral occipital cortex, and preferentially encoded words with properties that are associated with greater reliance on contextual or associative information, such as abstract words, words with sparse taxonomic structure, or words with dense associative semantic neighborhoods.

## Discussion

In this study, we identified latent dimensions of word-level structure during naturalistic narrative comprehension through a large-scale analysis of psycholinguistic properties spanning multiple theoretical domains. The eight latent factors capture core psycholinguistic constructs that reflect both classic and emerging theoretical distinctions in the literature. Each factor differentially accounted for behavioral performance across a range of psycholinguistic tasks, underscoring their cognitive relevance. Parcel-based univariate fMRI analyses revealed distinct neural activation patterns associated with each factor, while also identifying a set of partially overlapping regions, including the perisylvian language network (IFG/IFS and STG/STS extending to AG and SMG), left posterior MTG and vOTC, and right precuneus, that display hub-like properties by exhibiting sensitivity across multiple dimensions. These regions may therefore play integrative roles in supporting word processing during naturalistic comprehension. Beyond factor-specific localization, multivariate analyses of representational profiles further revealed four functionally distinct cortical subsystems. These subsystems were differentially associated with sensorimotor and salience-based representations, controlled semantic processing demands, lexical competition-related features, and contextual or associative semantics. Together, these findings suggest that multidimensional word representations during narrative comprehension are supported by interacting cortical systems distributed across fronto-temporal language regions and posterior medial cortex [22].

**A unified latent architecture underlies word representations**

Naturalistic language comprehension poses a fundamental modeling challenge: words vary simultaneously along a multitude of  of partially correlated psycholinguistic features, making it impractical and often conceptually incoherent to treat each variable as an independent predictor or control. Our large-scale factor analysis addresses this challenge by revealing a coherent latent architecture that captures the multifaceted properties of words. Across 106 psycholinguistic properties for approximately 14,000 English words, eight interpretable dimensions emerged. These dimensions encompassed usage-based properties (FA1 Frequency), form-based properties (FA2 phonological and orthographic neighborhood density and word length), bidirectional phonology–orthography mapping consistency (FA3 phonology-to-orthography consistency and FA4 orthography-to-phonology consistency), sublexical regularity (FA5 graphotactic and phonotactic probability), and multiple dimensions of semantic organization (FA6 Concreteness, FA7 taxonomic semantic neighborhood density, and FA8 associative semantic neighborhood density).

This latent structure provides a principled framework for interpreting relationships among psycholinguistic variables that are often treated as independent. For example, frequency, contextual diversity, age of acquisition, and semantic diversity are typically modeled as separable factors, yet they loaded strongly onto a single latent dimension. This pattern suggests that these measures share substantial variance that is consistent with cumulative language experience across contexts.

Similarly, the concreteness factor (FA6) integrated traditional concreteness ratings with imageability and multiple measures of sensory-related experience, unifying several constructs that are often examined in isolation. This factor showed the most extensive and robust neural modulation across cortical systems, consistent with the idea that sensorimotor features and experiential richness play a central role in organizing semantic representations during naturalistic language comprehension.

At the sublexical level, graphotactic and phonotactic probability captured shared variance among orthographic and phonological properties, underscoring the inherent difficulty of disentangling these features in alphabetic writing systems. At the semantic level, separating taxonomic and associative semantic neighborhood density revealed two dissociable dimensions that have frequently been conflated under the broader notion of semantic similarity[30,34,46]. Finally, the distinction between phonology-to-orthography and orthography-to-phonology consistency clarifies the bidirectional nature of phonology–orthography mapping in quasiregular writing systems such as English, and highlights the importance of operationalizing consistency at different sublexical grain sizes.

In contrast to dimensionality-reduction approaches such as PCA, which assumes orthogonality among components, the exploratory factor analysis used here allows latent factors to be intercorrelated, which better reflects the structure of inherently intercorrelated word properties. Consistent with this, several factors showed meaningful correlations, including FA1 Frequency with FA8 AssociativeSND, FA1 Frequency with FA2 PONeighborhood/Length, and FA2 PONeighborhood/Length with FA6 Concreteness. These relationships reflect the intrinsic dependencies among word properties and reinforce the need for modeling frameworks that preserve, rather than eliminate, such structure.

**The unified latent architecture predicts behavioral task performance**

All eight latent factors showed robust and interpretable relationships with behavioral performance across multiple psycholinguistic tasks, including auditory and visual lexical decision, word naming, word recognition, and semantic decision. This pattern indicates that each factor contributes to word processing beyond a single modality or processing stage. FA1 Frequency exerted the strongest and most consistent effects, with low-frequency words producing slower and less accurate responses, particularly in tasks requiring direct lexical access. Its influence was weaker in semantic decision after controlling for the effects of other factors, consistent with prior findings[42,47].

FA2 PONeighborhood/Length revealed a characteristic tradeoff [13,23,48–50]. Longer words with sparser phonological and orthographic neighborhoods were associated with slower and less accurate processing, reflecting increased decoding demands. In contrast, shorter words with denser neighborhoods likely incur greater lexical competition. The two consistency factors (FA3 POconsistency and FA4 OPconsistency) showed weaker and more task-dependent effects. In line with prior studies primarily using monosyllabic words[47,51–53], our results based on both monosyllabic and multisyllabic words showed that orthography-to-phonology consistency (FA4) facilitated performance in lexical decision and naming[28,54]. In contrast, phonology-to-orthography consistency (FA3) produced more selective and, in some cases, inhibitory effects. These results suggest that consistency influences processing in a manner that depends on directionality, grain size, and task demands[28,54,55]. Notably, these effects occur even in a purely auditory comprehension task, with no demands to either process orthography, or to connect it to phonology. This suggests a tight coupling of phonology and orthography, resulting in consistency effects even in the absence of reading.

FA5 Graphotactic/Phonotactic Probability further modulated behavioral performance across tasks. Higher scores on this factor were consistently associated with slower reaction times and reduced accuracy, suggesting an inhibitory influence of highly typical letter and sound sequences. This pattern is consistent with accounts in which increased similarity among sublexical neighbors heightens lexical competition, thereby slowing and destabilizing word recognition[50,56].

For the factors related semantic processing, words with higher concreteness scores (FA6) were processed more quickly and accurately, particularly in semantic decision and word recognition tasks, replicating the well-documented concreteness advantage in lexical processing[57–60]. Importantly, the two semantic neighborhood factors exhibited opposing behavioral effects[34,42]. Denser taxonomic neighborhoods interfered with processing, consistent with increased categorical competition. Denser associative neighborhoods facilitated performance, consistent with spreading activation and contextual support. Together, these patterns demonstrate that word processing is shaped by multiple interacting sources of facilitation and competition operating at lexical, sublexical, and semantic levels. The latent-factor framework provides a principled means of disentangling these effects.

**Distributed yet organized neural representations of lexical properties**

Rather than asking how individual psycholinguistic variables map onto the brain, we characterized the representational structure of word properties as a system and examined how this structure is reflected in neural activity during real-world language processing. Three organizing principles emerged from this analysis: (i) representations of word properties are distributed yet structured, (ii) word processing reflects a balance between integrative hubs and functionally specialized regions, and (iii) functional organization emerges in four subsystems at the level of coordinated representational profiles rather than gross anatomy alone.

First, representations of word-level dimensions were distributed across large-scale cortical systems yet exhibited clear and reproducible structure. Importantly, we modeled all latent dimensions simultaneously, which allowed us to disentangle neural effects that are often confounded in single-variable analyses. Each latent dimension showed a distinct neural footprint when modeled simultaneously with others (see Supplementary Discussion 1 for the discussion of each latent factor), but these footprints overlapped substantially within fronto-temporal language regions and posterior midline cortex. This organization contrasts with strictly modular accounts of language processing and instead supports a view in which psycholinguistic dimensions are encoded within overlapping representational spaces.

Second, the neural organization of word processing reflects a balance between integration and specialization. A set of regions, including inferior frontal cortex, superior and middle temporal cortex, posterior middle temporal gyrus, and precuneus, showed sensitivity across multiple latent dimensions, suggesting that they function as integrative hubs supporting flexible coordination of lexical, sublexical, and semantic information. In contrast, regions selectively modulated by one or two factors formed more distributed and functionally differentiated patterns. Semantic-related factors preferentially engaged regions including VMPFC, IFS/IFJ, TPOJ, PHG, and cingulate cortex, whereas form- and mapping-related factors showed greater selectivity in posterior FG and vOTC. Together, these findings indicate that word processing is supported by a combination of broadly integrative hubs and more specialized cortical territories.

Third, and critically, the pattern of overlap and selectivity revealed that functional organization cannot be fully captured at the level of individual regions. Parcels with similar representational profiles clustered together despite not always being spatially adjacent, whereas adjacent parcels could exhibit sharply distinct functional profiles. This dissociation highlights a network-level organization that is not reducible to gross anatomy alone. Multidimensional scaling further revealed systematic functional gradients across cortex, reflecting tradeoffs between abstract versus concrete representations, competition versus facilitation, and lexical versus contextual processing. Building on these representational profiles across the eight latent psycholinguistic dimensions, hierarchical clustering identified four interpretable subsystems that together support multidimensional word processing during narrative comprehension.

The first subsystem comprised bilateral SMG, TPOJ, PHG, MCC/dACC, POS, and left primary motor and somatosensory cortex. This system preferentially encoded words that are concrete, imageable, and grounded in sensorimotor or salient experience [14,16,45,61–66]. Its organization aligns with embodied and grounded theories of semantic representation [62,67], which posit that concrete concepts recruit perceptual, motor, and affective systems.

A second subsystem included bilateral IFS and IFJ, middle to anterior FG/ITG, left posterior MTG, and superior parietal lobule. This network was preferentially engaged by words that place greater demands on controlled semantic retrieval, such as low-frequency words, concrete words with numerous or competing semantic features, and words embedded in dense taxonomic neighborhoods. These regions closely correspond to the semantic control network identified in prior work single-word tasks that require effortful selection among competing representations and semantic aphasia patients who have brain damage particularly in those areas [68–70]. During narrative comprehension, this subsystem likely supports flexible access to meaning when automatic retrieval is insufficient.

The third subsystem encompassed bilateral DMPFC, MFG, IFG, STG, and STS.

Regions within this network were particularly sensitive to factors associated with lexical competition and phonology–orthography mapping, including dense phonological and orthographic neighborhoods and mapping inconsistency. The engagement of this subsystem during narrative listening suggests that, even in the absence of explicit metalinguistic demands, the language system continuously resolves competition among lexical candidates and sublexical mappings [61].

The fourth subsystem comprised bilateral PCC, precuneus, and lateral occipital cortex. This network preferentially encoded words that are more abstract, have sparser taxonomic neighborhoods, or have denser associative neighborhoods, indicating reliance on contextual, episodic, and integrative processes. These regions overlap with the default mode network, supporting interactions between semantic knowledge and episodic memory[61,71–73]. In narrative contexts, this subsystem likely facilitates the integration of word meaning with broader discourse structure, situational models, and internally generated associations[74,75]. Its engagement helps reconcile findings from narrative-listening studies that emphasize the role of posterior midline regions in meaning construction over extended timescales.

Together, these findings highlight word processing as a multidimensional, system-level phenomenon emerging from interactions among distributed cortical networks. The unified latent architecture introduced here offers a scalable and interpretable framework for studying language in ecologically valid contexts, where the complexity of word properties would otherwise be prohibitive to model directly.

More broadly, this approach reframes long-standing debates by shifting the focus from individual predictors to structured representational spaces, and by linking those spaces to neural systems operating during real-world language use. As such, it provides a foundation for future work examining individual differences, developmental trajectories, and clinical disruptions of language within a common representational coordinate system.

## Methods

### Psycholinguistic values from databases

We extracted 106 psycholinguistic variables from the SCOPE [42], encompassing general, phonological, orthographic, semantic, and morphological dimensions. Although SCOPE includes 105,992 unique word forms, only 5,553 have complete data across all selected variables. This limitation arises because many source databases provide values only for base forms (e.g., *abdicate* or *glass*), but not the inflected forms (e.g., *abdicated* and *abdicating*) or plural forms (e.g., *glasses*).

To maximize the number of tokens with available values for all selected variables, we implemented a three-step lemmatization procedure. First, we used the lemmas directly provided by the Subtlex-UK database [76], which employs Stanford NLP parsing based on the most frequent part-of-speech (PoS) tags. 61,842 words in SCOPE were lemmatized in this way. Second, for American spellings with available PoS tags provided in the *Subtlex-US* database but not the directly available dominated lemmas, we used the NLTK package [77] to extract the corresponding lemmas based on those most frequently used PoS tags. 5,717 words were lemmatized in this way. Third, for the remaining words ($N = 38,433$) that do not have available PoS tags, we used the natural language processing package spaCy [78] to automatically extract the lemma of every raw word form. As a result, our lemmatization pipeline resulted in a database consisting of 13,850 unique words (5905 unique lemmas) having available values for all the selected 106 variables. In the followed-up analysis, for the raw word forms whose psycholinguistic values were unavailable, we used the value of their lemmas as a best estimate. Additionally, for the variables of word frequency and contextual diversity, we used the lemma form measure summing across all word forms of the lemma.

**Exploratory factor analysis**

We used EFA to investigate the latent structure underlying the 106 psycholinguistic variables (KMO = 0.72). EFA was performed in python using the principal axis factoring extraction method and oblimin rotation. This method will search for the best latent structure while allowing for correlation between axes in higher dimension space. The number of factors to retain was determined based on the scree plot where the slope of the curve started to level off [43].

**Correlations between factor scores and behavioral outcome measures**

We leveraged the behavioral outcome variables provided in SCOPE [42] across five different tasks. Specifically, 15 outcome variables from seven major normative assessment megastudies were selected, including RT and ACC for visual lexical decision from the English Lexicon Project (ELP) [1], the British Lexicon Project (BLP) [4], and the English Crowdsourcing Project (ECP); RT and ACC for auditory lexical decision from the Auditory English Lexicon Project (AELP) [79] and the Massive Auditory Lexicl Decision database (MALD) [35]; RT and ACC for word naming from ELP; RT and ACC for concrete/abstract semantic decision from the Calgary database [33]; and the ACC for recognition memory performance indicated by hits minus false alarms [29]. The z-scored measures were used for all RT variables.

To examine the behavioral correlation of the identified factors, we calculated the bivariate Spearman correlation between the eight factor scores and behavioral outcomes. The sample sizes range from 2,426 to 11,626 words depending on databases. To obtain a robust estimation of the correlation between each factor and the overall performance for the same behavioral task across databases, we converted the correlation coefficients $r$ to Fisher's $z$, averaged the $z$ scores across databases for the same task, and then converted the averaged $z$ score back to $r$. This method has been shown to provide unbiased estimation when averaging correlation coefficients [80]. To reduce bias from large datasets, we used the harmonic mean of the sample sizes as the estimated sample size when converting back from $z$ to $r$ [81].

To evaluate the overall correlation between factor scores and behavioral outcomes, we also aggregated the correlation coefficients across three behavioral tasks (visual lexical decision, auditory lexical decision, and word naming) and four tasks (visual lexical decision, auditory lexical decision, word naming, and semantic decision), separately for RT and ACC. Additionally, the correlation coefficients were also aggregated across five tasks for ACC by adding the recognition memory task.

To determine statistical significance, we converted the correlation coefficients $r$ to Fisher's z and then to standard z. Considering that the correlation coefficients are prone to be statistically significant when sample size is large, we used a strigent threshold to determine significance ($p < .001$).

To control for the effect of other factors, we also calculated the bivariate Spearman partial correlation using the same pipeline.

**Comparison between factors and single variables**

We used a four-step pipeline to examine if the latent factors can outperform single variables in correlating with behavioral outcomes. First, for each behavioral outcome variable $Z$, we created a subset of words that have all available scores across the 106 psycholinguistic variables and $Z$. Seoncd, for each latent factor $Y$, we calculated the bivariate Spearman correlation $r_{YZ}$ between $Y$ and $Z$. Third, we calculated the Spearman correlation $r_{XZ}$ between $Z$ and each of the psycholinguistic variables that have high factor loadings (greater than 0.4) on factor $Y$, and identified the "best single predictor" $X$ that has the largest $r_{XZ}$ value. Then, we applied two-tailed dependent Steiger's T test [82] to compare if $r_{XZ}$ is significantly different from $r_{YZ}$ while account for the correlation $r_{XY}$ between the factor score and the "best single predictor". FDR correction was applied to account for multiple comparisons.

**fMRI data**

The fMRI data was leveraged from the "Narratives" data collection [44]. It contains 17 datasets of the fMRI recording while participants passively listening to different narratives in an MRI scanner. We selected 10 datasets that involved well-rehearsed narratives and provided any types of comprehension performance measures. Besides the participants that are recommended to be excluded based on the database guidebook, we also excluded the participants who had any type of the comprehension accuracy score less than 60% to ensure engagement during narrative listening. As a result, 213 fMRI scans from 102 neurotypical young adults ($M_{age} = 22.43$, $SD_{age} = 6.01$) were used.

All scans were acquired with the same repetition time (TR = 1500 ms) using three different ehco-planer imaging (EPI) sequences [44]. The same preprocessing pipeline was applied to all scans, including skull stripping, susceptibility distortion correction, realignment, co-registration, slice-timing correction, normalization, and resampling. The confounding time series were regressed out including the six head motion parameters (three translation and three rotation), the first five principal components from cerebrospinal fluid, the first five principal components from white matter, and the corresponding cosine bases for high-pass filtering under a discrete cosine filter with a 128s cut-off.

**fMRI data analysis**

*Parcel-wise analysis.* The images acquired with higher spatial resolution sequence were first resliced to match the images acquired with lower spatial resolution sequence (voxel size = 3×3×4 mm$^3$). For the first-level analysis, we used a parametric paradigm to investigate the modulation effect of psycholinguistic latent factors on BOLD signals using a general linear regression model (GLM) and the Finite Impulse Response (FIR) using AFNI's TENT function. The TENT function deconvolved the BOLD signal by modeling the hemodynamic response over a time window of 1.5 to 7.5 seconds after stimulus onset, using 5 tent functions to flexibly capture the shape of the hemodynamic function (HRF) without assuming a predefined form. The onset timestamps of content words (nouns, verbs, adjectives, adverbs, and pronouns, identified by the NLTK part-of-speech tagger [77]) were used as event timings in the model. The latent factor scores of content words were included as parametric modulators (using AFNI's AM2 setting). The onset timestamps of non-content words, as well as words that cannot be found in our factor analysis word database, were included as binary confounding variables in the model. Voxel-wise multiple linear regression was performed using 3dDeconvolve in AFNI. The signals in the primary auditory regions of the left hemisphere (left Heschl's gyri and parabelt areas) majorly involved in acoustic processing were regressed out. Considering that the oblimin rotation of factor analysis allows correlation among axes, we included all factors in the GLM to examine the modulation effect on the brain activity of a latent factor while controlling for the effects of other factors. Next, for each latent factor, we made a binary contrast by assigning 1 to the corresponding factor and zeros to all the other regressors, yielding eight whole-brain z-statistics contrast maps for the eight factors. Finally, we averaged the z-scores across the voxels within each of the 360 parcels from the HCP atlas, resulting eight parcel-based maps for each fMRI scan of each participant.

For the second-level analysis, for each factor, we applied one-sample t-tests for each parcel using 3dttest++ in AFNI, resulting in eight group-level parcel-based unthresholded z-statistics maps. Statistically significance was then determined at parcel-wise $p < .005$, with FDR correction applied for multiple comparisons.

*Localizing the parcels modulated by multiple factors.* To localize the brain regions that were broadly sensitive to multiple factors, we transformed the eight parcel-based group-level thresholded maps into eight binary maps and summed across maps to produce a frequency map of how many factors significantly modulated each parcel.

*Voxel-wise analysis.* As a validation analysis, we also examined the modulation effect of each factor on each brain voxel. To increase the signal-to-noise ratio, we used spatially smoothed the fMRI images by applying a Gaussian kernel of 6mm FWHM using 3dBlurToFWHM in AFNI. The first-level voxel-wise GLM was applied the same as that used for parcel-based analysis. For the second-level analysis, we used 3dttest++ in AFNI to examine the modulation effect on each voxel. The statistical significance of group-level maps was examined using a cluster-size based non-parametric thresholding method by using the 3dClustSim function in AFNI 3dttest++. This program estimates the noise structure in the data through randomization and permutation, and computes a null distribution of threshold cluster size that is needed to control for the false positive rate based on the provided brain masks. The clusters that are smaller than the threshold cluster size are considered statistically insignificant. We first examined our results with a graymatter mask to achieve a false positive rate at $\alpha$ = .05, voxel-wise $p$ < .005. To increase sensitivity in semantic processing and semantic control regions, we then estimated the cluster size using a predefined region of interest (ROI) encompassing canonical semantic processing and semantic control areas [68,72,83], including bilateral inferior frontal gyrus (IFG), angular gyrus (AG), precuneus, posterior cingulate cortex (PCC), fusiform, and parahippocampal gyrus (PHG). This ensures robust detection of both global and semantically specific effects.

*Multidimensional scaling analysis.* Following the parcel-wise parametric modulation analysis, we extracted eight group-level, parcel-based unthresholded $z$-statistic maps, corresponding to the eight latent psycholinguistic factors. For each cortical parcel, we constructed an 8-element feature vector comprising its modulation effects (group-level $z$-statistics) across the eight factors. Pairwise Euclidean distances between parcel vectors were then computed, yielding a symmetric parcel × parcel distance matrix $\mathbf{D}$, where each element is defined as $D_{ij} = \parallel v_i - v_j \parallel_2$.

Classical multidimensional scaling (MDS) was implemented to this distance matrix using the standard double-centering eigendecomposition procedure. Specifically, with $n$ parcels and centering matrix $H = I_n - \frac{1}{n}\mathbf{1}\mathbf{1}^\top$, we computed the Gram matrix $B = -\frac{1}{2}HD^2H$. Eigen decomposition of $B$ produced eigenvalues and eigenvectors which were sorted in descending order. Low-dimensional coordinates were obtained as $X = V\Lambda^{1/2}$, where $\mathbf{V}$ contains the selected eigenvectors and $\mathbf{\Lambda}$ is the diagonal matrix of corresponding eigenvalues. Variance explained by each dimension was computed as the proportion of each positive eigenvalue relative to the sum of all positive eigenvalues. We focused on the four leading MDS dimensions, which together accounted for 86.0% of the total variance.

For visualization, raw MDS coordinates along each component were min–max rescaled to the interval $[-1,1]$ and mapped to RGB colors using a continuous magenta–green color palette. Parcels that did not survive FDR correction in the parcel-wise analysis were masked from visualization.

*Hierarchical clustering analysis.* To further characterize the structure of representational similarity across cortical parcels, we performed hierarchical clustering on parcel-wise representational profiles derived from the parametric modulation analysis. Group-level, parcel-based unthresholded $z$-statistics were first reorganized into a parcel × factor feature matrix $\mathbf{X}$, such that each parcel was represented by an eight-dimensional vector corresponding to its modulation effects across the eight latent psycholinguistic factors.

Agglomerative hierarchical clustering was conducted based on $\mathbf{X}$ using Ward's variance-minimizing criterion, which iteratively merges parcels so as to minimize the increase in within-cluster variance under a Euclidean geometry. To improve dendrogram interpretability, the leaf ordering of the hierarchical tree was optimized to minimize distances between adjacent parcels while preserving the underlying hierarchical structure. The resulting dendrogram was partitioned into four clusters using a maximum-cluster criterion, yielding four major parcel groupings.

For surface-based visualization, cluster identities were mapped back onto the cortical surface. Each cluster was assigned a distinct categorical color, and parcels that did not survive FDR correction in the parcel-wise parametric modulation analysis were masked from visualization.

# Competing interests

The authors declare no competing interests.

# Acknowledgements

The research was supported by NIH/NIDCD R01DC017162 (RHD).

**Table 1.** Anatomical description of HCP cortical parcels

| HCP parcel | Gross anatomical location | Common anatomical label |
|---|---|---|
| Areas 9 | Superior frontal gyrus | Superior frontal gyrus |
| Areas 47 / 44 / 45 | Inferior frontal gyrus pars orbitalis, pars triangularis, and pars opercularis | Inferior Frontal Gyrus |
| IFSa / IFSp | Anterior and posterior portions of the inferior frontal sulcus and the inferior bank of middle frontal gyrus | Inferior frontal sulcus |
| IFJa / IFJp | Posterior portion of the inferior frontal sulcus | Inferior frontal sulcus/inferior frontal junction |
| SFL | Medial superior frontal gyrus, dorsal to the cingulate sulcus | Superior frontal language area; Supplemental motor area |
| Area 4 | Posterior half of the precentral gyrus and the anterior bank of the central sulcus | Precentral gyrus; primary motor cortex |
| Area 3a | In the depth of central sulcus | Central sulcus; primary sensory area |
| Area 3b | Anterior bank of the postcentral gyrus | Posterior central gyrus; primary sensory area |
| FEF | Anterior half of the precentral gyrus, approximately half way down its length along the convexity, inferior to the junction point of the precentral and superior frontal sulci | Precentral gyrus |
| 55b | Anterior half of the precentral gyrus extending to posterior edge of the middle frontal gyrus | Precentral gyrus/posterior middle frontal gyrus |
| PEF | The floor of the precentral sulcus at the junction of the precentral and inferior frontal sulci | Precentral gyrus |
| Areas 24 / 32 / 25 / 33 | Different subregions of anterior cingulate gyrus | Anterior cingulate cortex |
| Areas AAIC, AVI, MI, FOP | Different subregions of insula and opercular cortex | Insula/opercular cortex |
| A4 / A5 | Auditory association cortex | Superior temporal gyrus |
| STSda / STSdp | Dorsal bank of the superior temporal sulcus | Superior temporal sulcus (dorsal) |
| STSva / STSvp | Ventral bank of the superior temporal sulcus | Superior temporal sulcus (ventral) |
| TE1a / TE1m / TE1p | Anterior, middle, and posterior portions of middle temporal gyrus | Middle temporal gyrus |
| TE2a / TE2p | Anterior and posterior portion of the inferior temporal gyrus | Inferior temporal gyrus |
| TGd / TGv | Superior and inferior part of the temporopolar region | Temporal pole/anterior temporal lobe |

| TF | Anterior part of the fusiform gyrus and the occipitotemporal sulcus | Fusiform gyrus/Inferior temporal gyrus |
|---|---|---|
| TGd/TGv | Superior and inferior portion of the temporal polar region just anterior to the ITG and fusiform gyrus | Temporal Pole |
| PreS | Posterior superior surface of the parahippocampal gyrus | Parahippocampal gyrus |
| Areas PHA | Medial temporal lobe along the parahippocampal gyrus | Parahippocampal cortex |
| PFop | Anterior portion of the inferior parietal lobule | Inferior parietal Lobule |
| PFt | Posterior bank of the postcentral sulcus, on the anterior superior edge of the inferior parietal lobule | Inferior parietal Lobule |
| PF | Lateral surface of the superior portion of the supramarginal gyrus | Supramarginal gyrus |
| PHT | posterior middle temporal gyrus leading into the angular gyrus | pMTG/AG |
| PFm | Anterior superior surface of the angular gyrus and the posterior superior bank of the supramarginal gyrus | Angular gyrus/Supramarginal gyrus |
| PGs / PGi | Superior and inferior surface of the angular gyrus | Angular gyrus |
| PGp | The most posterior portion of the inferior parietal lobule | Inferior parietal lobule |
| Areas TPOJ | Junction of temporal, parietal, and occipital lobes | Temporo-parieto-occipital junction |
| Areas IPS | Intraparietal sulcus | Intraparietal sulcus |
| | | |
| Areas 31 | Different subregions of subparietal area | Subparietal gyrus/posterior cingulate cortex |
| Areas 23 | Posterior cingulate gyrus | Posterior cingulate cortex |
| RSC | A long thin area of posterior cingulate cortex that is immediately adjacent to the colossal sulcus wrapping around the splenium | Retrosplenial cortex; posterior cingulate cortex |
| Areas POS | Different subregions of parieto-occipital sulcus | Parieto-occipital sulcus |
| PCV | Precuneus visual area/anterior precuneus | Precuneus |
| 7m | Posterior precuneus | Precuneus |
| V1-V4 | Primary visual areas | Primary visual cortex |
| V2 | Medial and lateral occipital cortex immediately surrounding V1 | Secondary visual cortex |
| V3 / V4 | Medial and lateral occipital cortex anterior to V2 | Extrastriate visual cortex |
| Areas LO | Lateral surface of the occipital lobe | Lateral occipital cortex |

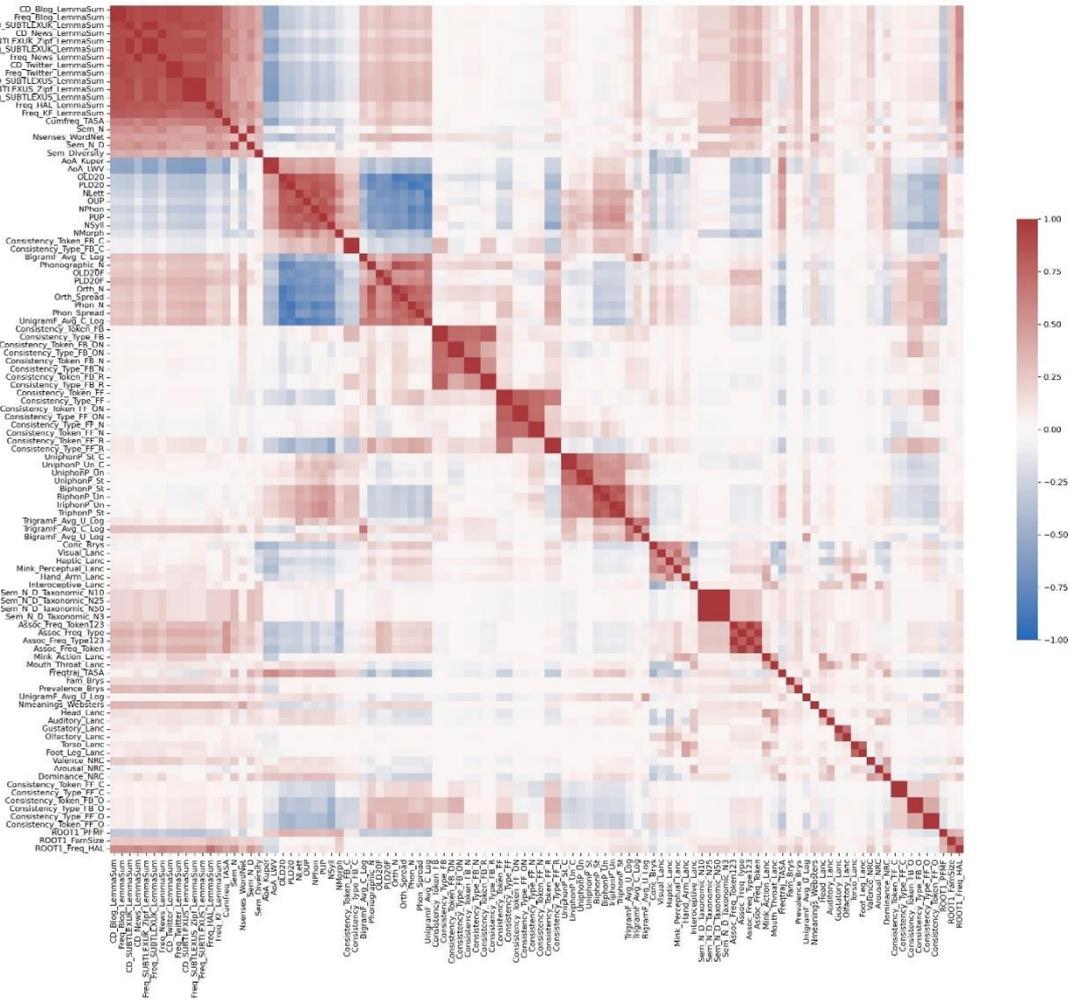

*Fig. 1 | Spearman correlations between the 106 selected variables suggested latent structures*

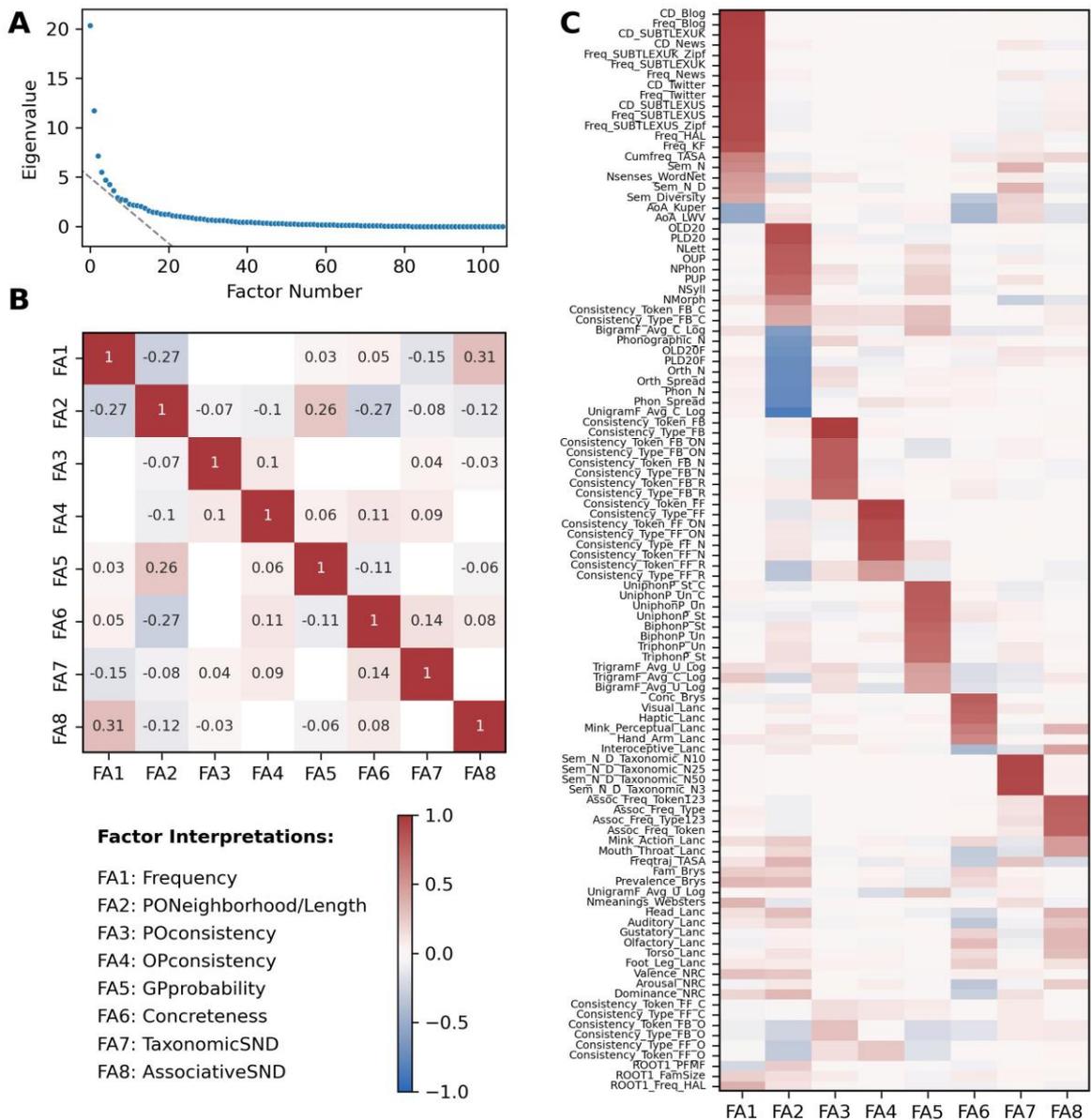

**Fig. 2 | EFA suggested eight interpretable latent factors.**
**A.** The scree plot of factor analysis. The dashed line indicates the elbow at eight components. **B.** The pairwise Spearman correlations that were statistically significant, $p < .001$. **C.** The heatmap of factor loadings for the eight extracted factors.

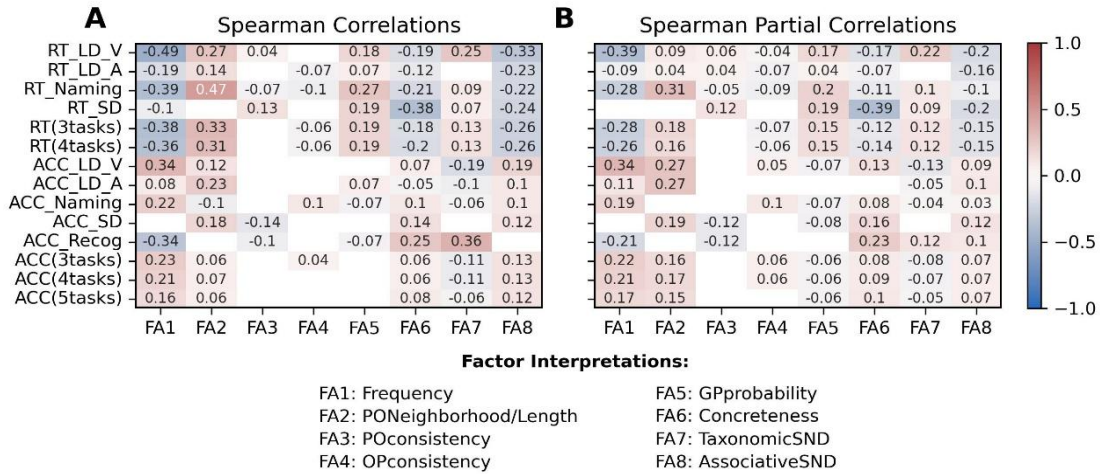

**Fig. 3 | The correlations between factor scores and behavioral outcome measures.**
**A.** The pairwise Spearman correlations between factor scores and reaction times and accuracy in different tasks. **B.** The pairwise partial Spearman correlations between each factor score and reaction times and accuracy after controlling for the other factors. *Notes.* Only significant correlations were shown in the figure, $p < .001$. LD_V: visual lexical decision. LD_A: auditory lexical decision. Naming: word naming. SD: semantic decision. RT (3tasks): the weighted average for reaction times across LD_V, LD_A, and Naming. RT (4tasks): the weighted average for reaction times across LD_V, LD_A, Naming, and SD. ACC (3tasks): the weighted average for accuracy across LD_V, LD_A, and Naming. ACC (4tasks): the weighted average for accuracy across LD_V, LD_A, Naming, and SD. ACC(5tasks): the weighted average for accuracy across LD_V, LD_A, Naming, SD, and Recog.

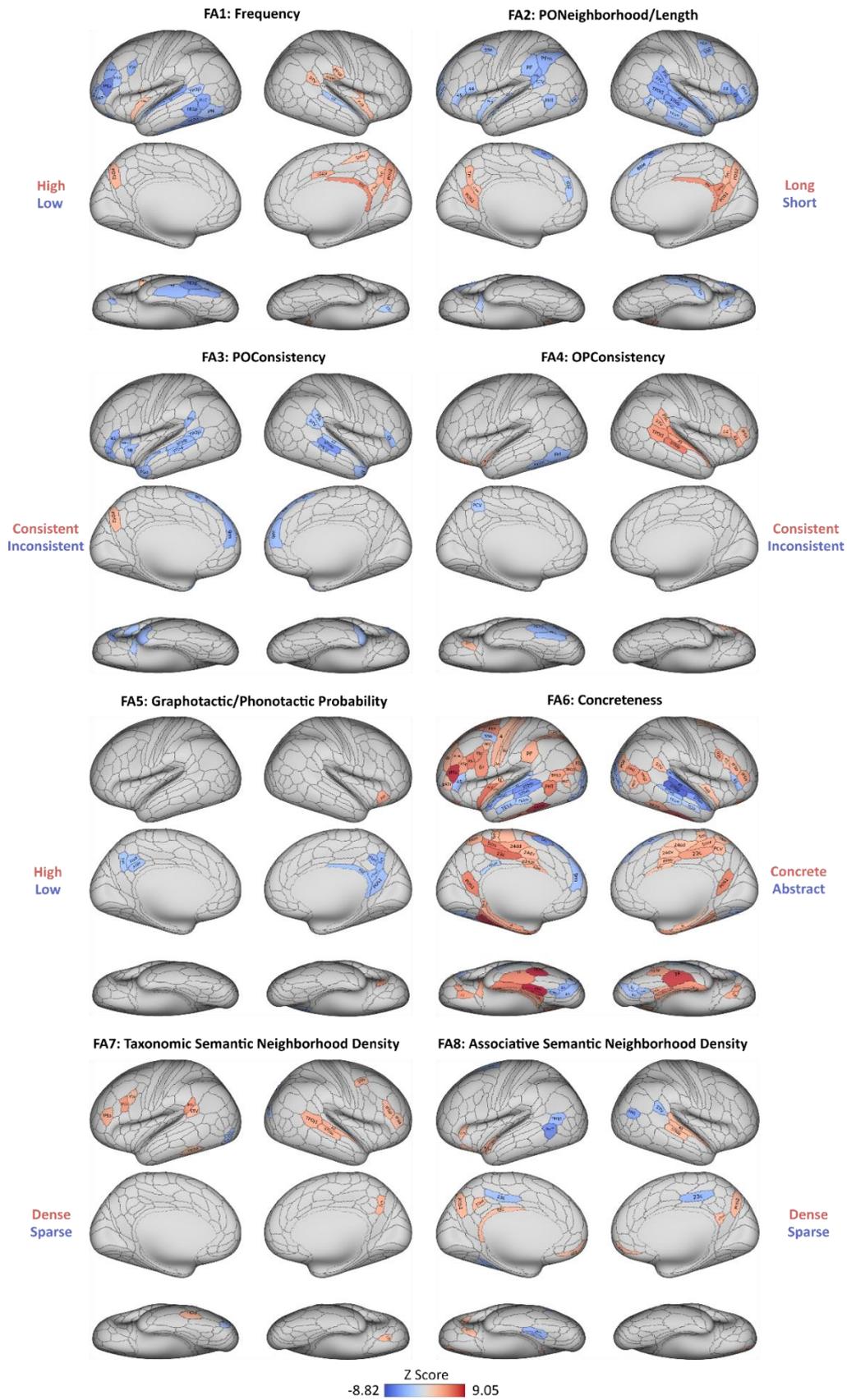

**Fig. 4 | Parcel-wise modulation maps showing how BOLD was modulated by each latent factor after controlling for the other factors.**

*Notes*. Only statistically significant results were shown in the figure. *p* < .005, FDR corrected.

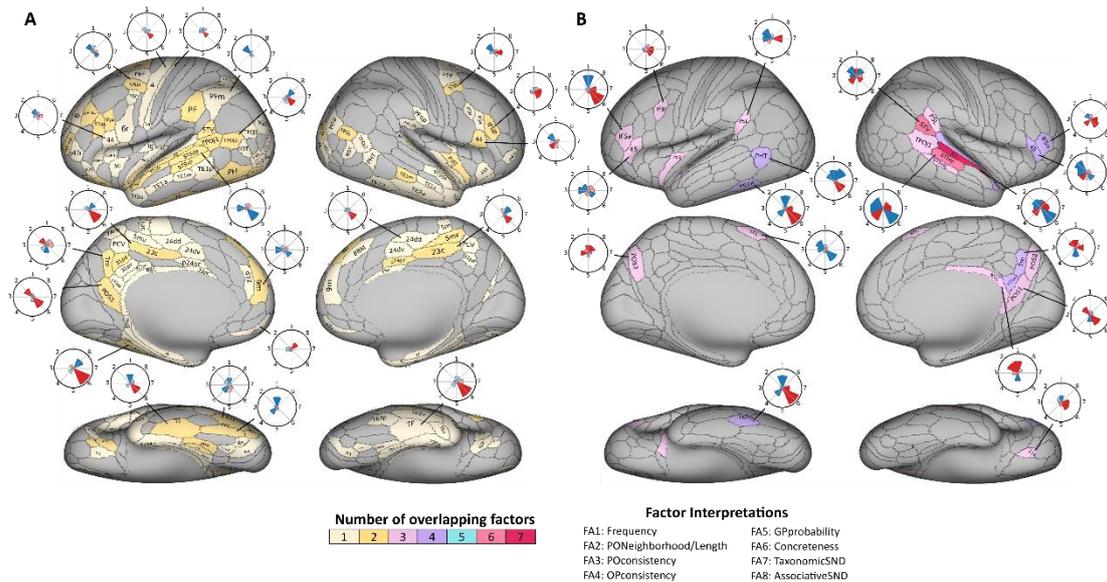

**Fig. 5 | Cortical parcels grouped by the number of latent factors modulating their activity.**

**A.** Regions modulated by fewer than three latent factors. **B.** Regions modulated by three or more latent factors. For selected cortical parcels, representational profiles of the eight latent factors are visualized using circular bar plots. Each sector corresponds to one latent factor, with radial length proportional to the unique modulation effect of that factor while controlling for the others. Sector fill color indicates statistical significance (red = significantly positive, blue = significantly negative, gray = non-significant; $p < 0.005$, FDR-corrected). The edge color of each sector indicates the sign of modulation.

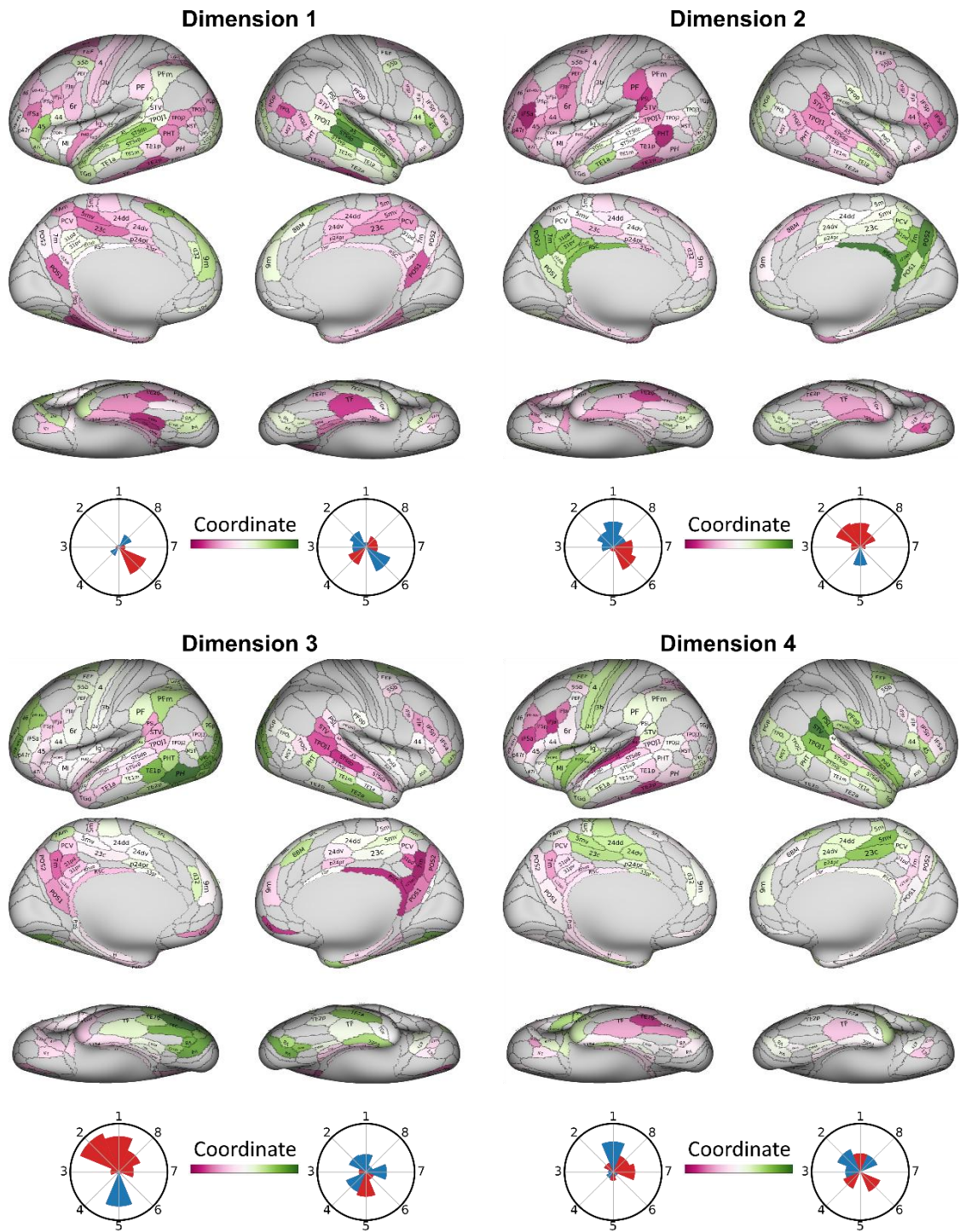

**Fig 6. | MDS suggests distinct neural representation patterns underlying word processing during narrative listening.**

Notes. For visualization, raw MDS coordinates along each component were min–max rescaled to the interval $[-1,1]$ and mapped to RGB colors using a continuous magenta–green color palette. Parcels that did not survive FDR correction in the parcel-wise analysis were masked from visualization. The circular bar plots show the averaged representational profiles of the top three parcels at each end of the dimension.

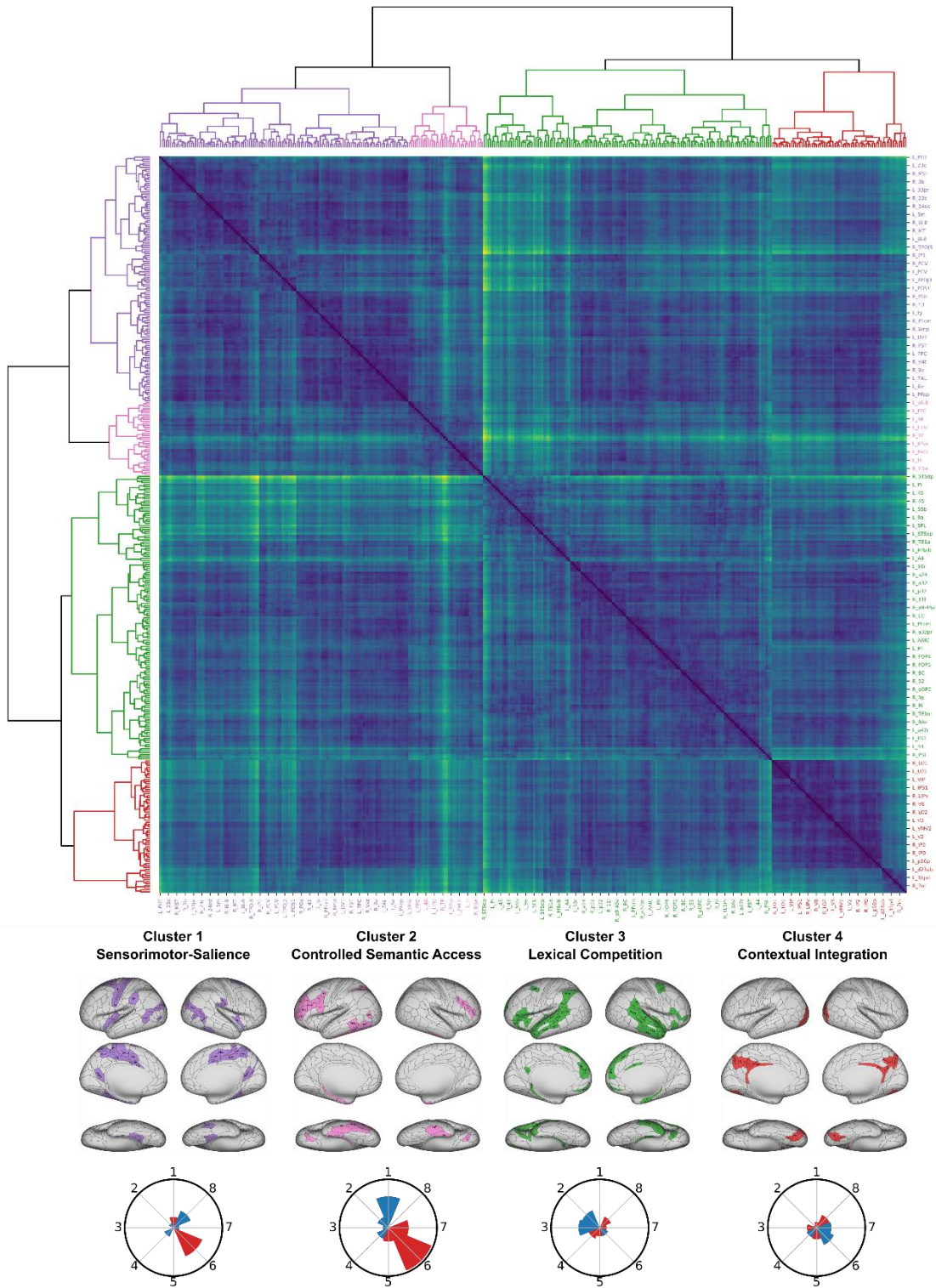

**Fig 7. | Hierarchical cluster analysis identified four distinct subsystems underlying word processing during narrative listening.**

Hierarchical clustering was applied to parcel-wise representational profiles derived from the parametric modulation analysis. The resulting representational similarity matrix (top) is shown with parcels reordered according to the optimized dendrogram leaf ordering, which minimizes distances between adjacent parcels while preserving the hierarchical structure. The dendrogram was partitioned into four clusters using a maximum-cluster criterion, yielding four major parcel groupings (indicated by color). For surface-based visualization (middle), cluster identities were mapped back onto the cortical surface and displayed using categorical colors. Parcels that did not survive FDR correction in the parcel-wise parametric modulation analysis were masked from visualization. Polar bar plots (bottom) depict the averaged representational profiles across all parcels within each cluster, illustrating cluster-level modulation patterns across the eight psycholinguistic factors.

# Supplemental materials

## Supplementary Discussion

### Unique neural activity patterns for latent dimensions

Examining how latent dimensions of word properties are represented in the brain during narrative listening provides an ecologically valid perspective on language processing and helps reconcile findings from traditional single-word paradigms[1–5]. By modeling all eight factors simultaneously, we were able to dissociate neural systems that uniquely track specific psycholinguistic dimensions while also identifying regions that integrate multiple sources of lexical, sublexical, and semantic information. Parcel-based analyses revealed that although each factor exhibited a distinct spatial pattern, their neural representations partially converged in distributed fronto-temporo-parietal and posterior midline cortical regions, consistent with integrative cortical systems supporting speech, semantics, and discourse-level processing[4,6–10].

The neural mechanisms underlying frequency effects remain debated, with prior accounts emphasizing either increased demands on phonology–orthography mapping or greater lexicosemantic processing for low-frequency words, and enhanced semantic access for high-frequency words[1,11–15]. In the present study, after controlling for other psycholinguistic dimensions, low-frequency words were associated with increased activation in left STG, IFS and IFJ, as well as left FG/vOTC. Conjunction analyses indicated that these regions overlapped with factors related to both semantic processing and phonology–orthography mapping, suggesting that frequency-related modulation reflects a combination of increased lexicosemantic demands and sublexical mapping processes. In contrast, higher-frequency words were associated with increased activation in right posterior STG, SMG, PCC, and precuneus, consistent with facilitated semantic access and greater engagement of memory- and context-related systems during naturalistic comprehension.

Higher FA2 scores, corresponding to longer words with sparser phonological and orthographic neighborhoods, were associated with reduced activation in bilateral IFG, SFG, MFG, posterior STG, SMG, and AG, alongside increased activation in PCC and precuneus. This pattern is in line with prior work has similarly reported broad deactivations for composite measures of word length and neighborhood size during passive reading and listening tasks[12,16]. In passive listening contexts, shorter words with dense neighborhoods likely elicit stronger lexical competition, thereby increasing recruitment of frontal and temporal regions involved in selection and control. In contrast, longer words with sparser neighborhoods may rely more on contextual integration and episodic processing, reflected in increased posterior midline activation.

### Phonology–orthography consistency

The two consistency factors revealed partially overlapping but dissociable neural effects. Regions including IFG, STG, and STS were sensitive to mapping consistency in both directions, particularly under conditions of increased competition from other lexical and semantic sources. However, posterior FG and vOTC showed selective sensitivity to orthography-to-phonology inconsistency, consistent with its established role in grapheme–phoneme conversion and feedforward mapping[17–20]. These findings demonstrate that phonology–orthography mappings influence spoken language processing even in the absence of orthographic input. Feedback inconsistency produced activation in the anterior and middle temporal lobe. This meshes well with studies of patients with primary progressive aphasia showing that temporal atrophy typically results in surface dyslexia, which is characterized by deficits in reading inconsistent words[21,22].

**Graphotactic and phonotactic probability**

Neural modulation by FA5 (graphotactic/phonotactic probability) during narrative listening was relatively weaker and more spatially restricted compared with other latent factors, suggesting that this sublexical property plays a more limited role in dynamic word processing under naturalistic listening conditions. Parcel-wise analysis identified only a single region, the anterior ventral insula (parcel AVI), showing a significant positive association with this factor. Complementary voxel-wise analyses revealed positive correlations in bilateral inferior frontal gyrus pars triangularis, inferior parietal lobule, and superior frontal gyrus—regions implicated in semantic selection and domain-general cognitive control[23–26]. This neural pattern aligns with the behavioral finding that higher FA5 scores were associated with longer reaction times across tasks, consistent with increased processing demands when letter or sound sequences are less predictable. In addition, we observed negative correlations in PCC and precuneus, indicating greater activation in these posterior midline regions for words with lower graphotactic or phonotactic probability. Given the involvement of PCC and precuneus in episodic, contextual, and internally oriented processing, one conservative interpretation is that atypical sublexical structure increases reliance on contextual or memory-based mechanisms during word recognition.

**Concreteness**

The Concreteness factor elicited broad and distributed modulation across both sensory and integrative semantic systems. Across analyses, it emerged as the dominant contributor to the first dimension of the multidimensional scaling solution and formed the core of the first cluster in the hierarchical clustering analysis. Together, these results indicate that concreteness captures a major axis of variation in how words and concepts are represented in the brain. This pattern aligns with prior evidence of semantic processing and supports a view in which concrete concepts are encoded through the joint engagement of unimodal sensory–motor regions, such as premotor cortex and higher-order visual and auditory areas, as well as multimodal integrative regions including AG and PHG [2,7,27–29]. In contrast, abstract words were associated with greater activation in bilateral IFG, STG, STS, SFG, and MFG. These regions substantially overlap with the subsystem implicated in resolving lexical competition and controlled access to meaning[30,31], consistent with the our behavioral finding that abstract words are processed more slowly and less accurately. In addition, increased activation in bilateral STG, dorsomedial prefrontal cortex, and lateral occipital cortex suggests that abstract concepts may also be grounded in semantic features such as emotion and social relationships[4,32,33]. This interpretation is consistent with the established roles of superior temporal, medial prefrontal, and lateral occipital regions in emotional and social–conceptual processing[34,35]. Compared with prior neuroimaging evidence based primarily on single-word paradigms, the present findings revealed more extensive and bilateral effects, likely reflecting the richer semantic integration and contextual demands inherent to narrative comprehension[4].

**Taxonomic and associative semantic neighborhood density**

Taxonomic and associative semantic neighborhood density showed both shared and distinct neural representations. Denser neighborhoods of both types were associated with increased activation in bilateral IFG, IFJ, pSTG, STS, and precuneus, implicating a common semantic control and integration network[7,26,36]. However, associative neighborhood density uniquely engaged VMPFC and left anterior STG, suggesting a role for affective and thematic associations in accessing associative semantic structure[37–39]. In contrast, sparser neighborhoods elicited greater activation in distinct systems. Sparse taxonomic neighborhoods were associated with increased activation in LOC and SPL, consistent with greater reliance on perceptual and spatial features when categorical structure is weak. Sparse associative neighborhoods were linked to increased activation in PHG and pMTG, indicating greater dependence on episodic and contextual retrieval when associative links are limited[27]. Together, these findings highlight a qualitative distinction between categorical similarity and thematic association, supported by partially overlapping but functionally differentiated neural systems.

**Table S1: Source of databases**

| Level 1 | Level 2 | Name | Source | Name in Source Database | Citation(Full) |
|---|---|---|---|---|---|
| General | Frequency | Freq_HAL | ELP | Log_Freq_HAL | Lund, K. and C. Burgess (1996). "Producing high-dimensional semantic spaces from lexical co-occurrence." Behavior Research Methods, Instruments, & Computers 28(2): 203-208. |
| General | Frequency | Freq_KF | ELP | Freq_KF | Kučera, H. and W. N. Francis (1967). Computational analysis of present-day American English, Brown University Press. |
| General | Frequency | Freq_SUBTL EXUS | SUBTLEX-US | Lg10WF | Brysbaert, M. and B. New (2009). "Moving beyond Kučera and Francis: A critical evaluation of current word frequency norms and the introduction of a new and improved word frequency measure for American English." Behavior Research Methods41(4): 977-990. |
| General | Frequency | Freq_SUBTL EXUS_Zipf | SUBTLEX-US | Zipf-value | Brysbaert, M. and B. New (2009). "Moving beyond Kučera and Francis: A critical evaluation of current word frequency norms and the introduction of a new and improved word frequency measure for American English." Behavior Research Methods41(4): 977-990. Van Heuven, W. J., et al. (2014). "SUBTLEX-UK: A new and improved word frequency database for British English." Quarterly Journal of Experimental Psychology 67(6): 1176-1190. |
| eneral | Frequency | Freq_SUBTL EXUK | SUBTLEX-UK | FreqCount (Log10 of this variable is reported) | Van Heuven, W. J., et al. (2014). "SUBTLEX-UK: A new and improved word frequency database for British English." Quarterly Journal of Experimental Psychology67(6): 1176-1190. |
| General | Frequency | Freq_SUBTL EXUK_Zipf | SUBTLEX-UK | LogFreq(Zipf) | Van Heuven, W. J., et al. (2014). "SUBTLEX-UK: A new and improved word frequency database for British English." Quarterly Journal of Experimental Psychology67(6): 1176-1190. |
| General | Frequency | Freq_Blog | Worldlex | BlogFreq | Gimenes, M. and B. New (2016). "Worldlex: Twitter and blog word frequencies for 66 languages." Behavior Research Methods48(3): 963-972. |
| General | Frequency | Freq_Twitter | Worldlex | TwitterFreq | Gimenes, M. and B. New (2016). "Worldlex: Twitter and blog word frequencies for 66 languages." Behavior Research Methods48(3): 963-972. |
| General | Frequency | Freq_News | Worldlex | NewsFreq | Gimenes, M. and B. New (2016). "Worldlex: Twitter and blog word frequencies for 66 languages." Behavior Research Methods48(3): 963-972. |

| General | Contextual Diversity | CD_SUBTLEXUS | SUBTLEX-US | Lg10CD | Brysbaert, M. and B. New (2009). "Moving beyond Kučera and Francis: A critical evaluation of current word frequency norms and the introduction of a new and improved word frequency measure for American English." Behavior Research Methods41(4): 977-990. |
|---------|---------|---------|---------|---------|---------|
| General | Contextual Diversity | CD_SUBTLEXUK | SUBTLEX-UK | CD (Log10 of this variable is reported) | Van Heuven, W. J., et al. (2014). "SUBTLEX-UK: A new and improved word frequency database for British English." Quarterly Journal of Experimental Psychology67(6): 1176-1190. |
| General | Contextual Diversity | CD_Blog | Worldlex | BlogCD | Gimenes, M. and B. New (2016). "Worldlex: Twitter and blog word frequencies for 66 languages." Behavior Research Methods48(3): 963-972. |
| General | Contextual Diversity | CD_Twitter | Worldlex | TwitterCD | Gimenes, M. and B. New (2016). "Worldlex: Twitter and blog word frequencies for 66 languages." Behavior Research Methods48(3): 963-972. |
| General | Contextual Diversity | CD_News | Worldlex | NewsCD | Gimenes, M. and B. New (2016). "Worldlex: Twitter and blog word frequencies for 66 languages." Behavior Research Methods48(3): 963-972. |
| General | Familiarity | Fam_Brys | Brysbaert14 | Percent_known | Brysbaert, M., et al. (2014). "Concreteness ratings for 40 thousand generally known English word lemmas." Behavior Research Methods46(3): 904-911. |
| General | Familiarity | Prevalence_Brys | Brysbaert19 | Prevalence | Brysbaert, M., et al. (2019). "Word prevalence norms for 62,000 English lemmas." Behavior Research Methods51(2): 467-479. |
| General | Age of Acquisition | AoA_Kuper | Kuperman12 | Age_Of_Acquisition | Kuperman, V., et al. (2012). "Age-of-acquisition ratings for 30,000 English words." Behavior Research Methods44(4): 978-990. |
| General | Age of Acquisition | AoA_LWV | Brysbaert17 | AoA | Brysbaert, M. (2017). "Age of acquisition ratings score better on criterion validity than frequency trajectory or ratings "corrected" for frequency." Quarterly Journal of Experimental Psychology70(7): 1129-1139. Dale, E. and J. O'Rourke (1981). The Living Word Vocabulary. Chicago: World Book-Childcraft International, Inc. |
| General | Frequency Trajectory | Freqtraj_TASA | Brysbaert17 | Freqtraj | Brysbaert, M. (2017). "Age of acquisition ratings score better on criterion validity than frequency trajectory or ratings "corrected" for frequency." Quarterly Journal of Experimental Psychology70(7): 1129-1139. |

| General | Frequency Trajactory | Cumfreq_TASA | Brysbaert17 | Cumfreq | Brysbaert, M. (2017). "Age of acquisition ratings score better on criterion validity than frequency trajectory or ratings "corrected" for frequency." Quarterly Journal of Experimental Psychology70(7): 1129-1139. |
|---|---|---|---|---|---|
| Orthographic | Orthographic Length | NLett | | | |
| Orthographic | Graphotactic Probabilities | UnigramF_Avg_C_Log | MCWord | N1_F | Medler, D. and J. Binder (2005). MCWord: An on-line orthographic database of the English language. |
| Orthographic | Graphotactic Probabilities | UnigramF_Avg_U_Log | MCWord | UN1_F | Medler, D. and J. Binder (2005). MCWord: An on-line orthographic database of the English language. |
| Orthographic | Graphotactic Probabilities | BigramF_Avg_C_Log | MCWord | N2_F | Medler, D. and J. Binder (2005). MCWord: An on-line orthographic database of the English language. |
| Orthographic | Graphotactic Probabilities | BigramF_Avg_U_Log | MCWord | UN2_F | Medler, D. and J. Binder (2005). MCWord: An on-line orthographic database of the English language. |
| Orthographic | Graphotactic Probabilities | TrigramF_Avg_C_Log | MCWord | N3_F | Medler, D. and J. Binder (2005). MCWord: An on-line orthographic database of the English language. |
| Orthographic | Graphotactic Probabilities | TrigramF_Avg_U_Log | MCWord | UN3_F | Medler, D. and J. Binder (2005). MCWord: An on-line orthographic database of the English language. |
| Orthographic | Orthographic Neighborhood | OLD20 | ELP | OLD | Yarkoni, T., et al. (2008). "Moving beyond Coltheart's N: A new measure of orthographic similarity." Psychonomic bulletin & review15(5): 971-979. |
| Orthographic | Orthographic Neighborhood | OLD20F | ELP | OLDF | Yarkoni, T., et al. (2008). "Moving beyond Coltheart's N: A new measure of orthographic similarity." Psychonomic bulletin & review15(5): 971-979. |
| Orthographic | Orthographic Neighborhood | Orth_N | ELP | Ortho_N/Coltheart's N | Balota, D. A., et al. (2007). "The English lexicon project." Behavior Research Methods39(3): 445-459. Coltheart, M. (1977). "Access to the internal lexicon." The psychology of reading. |

| Orthographic | Orthographic Neighborhood | Orth_Spread | Chee20 | Ortho_spread | Chee, Q. W., et al. (2020). "Consistency norms for 37,677 english words." Behavior Research Methods. |
|---|---|---|---|---|---|
| Orthographic | Orthographic Neighborhood | OUP | MALD | OUP | Tucker, B. V., et al. (2019). "The massive auditory lexical decision (MALD) database." Behavior Research Methods51(3): 1187-1204. Weide, R. (2005). "The Carnegie mellon pronouncing dictionary [cmudict. 0.6]." Pittsburgh, PA: Carnegie Mellon University. |
| Phonological | Phonological Length | NPhon | ELP | NPhon | Balota, D. A., et al. (2007). "The English lexicon project." Behavior Research Methods39(3): 445-459. |
| Phonological | Phonological Length | NSyll | ELP | NSyll | Balota, D. A., et al. (2007). "The English lexicon project." Behavior Research Methods39(3): 445-459. |
| Phonological | Phonotactic Probabilities | UniphonP_Un | IPhoD | unsLPOSPAV | Vaden, K. I., et al. (2009). Irvine phonotactic online dictionary, Version 2.0.[Data file]. |
| Phonological | Phonotactic Probabilities | UniphonP_St | IPhoD | strLPOSPAV | Vaden, K. I., et al. (2009). Irvine phonotactic online dictionary, Version 2.0.[Data file]. |
| Phonological | Phonotactic Probabilities | UniphonP_Un_C | IPhoD | unsLLCPOSPAV | Vaden, K. I., et al. (2009). Irvine phonotactic online dictionary, Version 2.0.[Data file]. |
| Phonological | Phonotactic Probabilities | UniphonP_St_C | IPhoD | strLLCPOSPAV | Vaden, K. I., et al. (2009). Irvine phonotactic online dictionary, Version 2.0.[Data file]. |
| Phonological | Phonotactic Probabilities | BiphonP_Un | IPhoD | unsLBPAV | Vaden, K. I., et al. (2009). Irvine phonotactic online dictionary, Version 2.0.[Data file]. |
| Phonological | Phonotactic Probabilities | BiphonP_St | IPhoD | strLBPAV | Vaden, K. I., et al. (2009). Irvine phonotactic online dictionary, Version 2.0.[Data file]. |
| Phonological | Phonotactic Probabilities | TriphonP_Un | IPhoD | unsLTPAV | Vaden, K. I., et al. (2009). Irvine phonotactic online dictionary, Version 2.0.[Data file]. |
| Phonological | Phonotactic Probabilities | TriphonP_St | IPhoD | strLTPAV | Vaden, K. I., et al. (2009). Irvine phonotactic online dictionary, Version 2.0.[Data file]. |
| Phonological | Phonological Neighborhood | PLD20 | ELP | PLD | Balota, D. A., et al. (2007). "The English lexicon project." Behavior Research Methods39(3): 445-459. |
| Phonological | Phonological Neighborhood | PLD20F | ELP | PLDF | Balota, D. A., et al. (2007). "The English lexicon project." Behavior Research Methods39(3): 445-459. |
| Phonological | Phonological Neighborhood | Phon_N | ELP | Phono_N | Balota, D. A., et al. (2007). "The English lexicon project." Behavior Research Methods39(3): 445-459. |
| Phonological | Phonological Neighborhood | Phon_Spread | Chee20 | Phono_spread | Chee, Q. W., et al. (2020). "Consistency norms for 37,677 english words." Behavior Research Methods. |
| Phonological | Phonological Neighborhood | PUP | MALD | PUP | Tucker, B. V., et al. (2019). "The massive auditory lexical decision (MALD) database." Behavior Research Methods51(3): 1187-1204. Weide, R. (2005). "The Carnegie mellon pronouncing dictionary [cmudict. 0.6]." Pittsburgh, PA: Carnegie Mellon University. |
| Semantic | Concreteness/ Imageability | Conc_Brys | Brysbaert14 | Conc.M | Brysbaert, M., et al. (2014). "Concreteness ratings for 40 thousand generally known English word lemmas." Behavior Research Methods46(3): 904-911. |
| Semantic | Polysemy | Nsenses_WordNet | WordNet | Nsenses | Miller, G. A. (1995). "WordNet: a lexical database for English." Communications of the ACM38(11): 39-41. |
| Semantic | Polysemy | Nmeanings_Websters | Websters | Number of meanings | Gao, C., Shinkareva, S. V., & Desai, R. H. "SCOPE: The South Carolina Psycholinguistic Metabase" (forthcoming) |
| Semantic | Specific Semantic Features | Visual_Lanc | Lancaster | Visual | Lynott, D., et al. (2020). "The Lancaster Sensorimotor Norms: Multidimensional measures of Perceptual and Action Strength for 40,000 English words." Behavior Research Methods52(3): 1271-1291. |

| Semantic | Specific Semantic Features | Auditory_Lanc | Lancaster | Auditory | Lynott, D., et al. (2020). "The Lancaster Sensorimotor Norms: Multidimensional measures of Perceptual and Action Strength for 40,000 English words." Behavior Research Methods52(3): 1271-1291. |
|---|---|---|---|---|---|
| Semantic | Specific Semantic Features | Haptic_Lanc | Lancaster | Haptic | Lynott, D., et al. (2020). "The Lancaster Sensorimotor Norms: Multidimensional measures of Perceptual and Action Strength for 40,000 English words." Behavior Research Methods52(3): 1271-1291. |
| Semantic | Specific Semantic Features | Olfactory_Lanc | Lancaster | Olfactory | Lynott, D., et al. (2020). "The Lancaster Sensorimotor Norms: Multidimensional measures of Perceptual and Action Strength for 40,000 English words." Behavior Research Methods52(3): 1271-1291. |
| Semantic | Specific Semantic Features | Gustatory_Lanc | Lancaster | Gustatory | Lynott, D., et al. (2020). "The Lancaster Sensorimotor Norms: Multidimensional measures of Perceptual and Action Strength for 40,000 English words." Behavior Research Methods52(3): 1271-1291. |
| Semantic | Specific Semantic Features | Interoceptive_Lanc | Lancaster | Interoceptive | Lynott, D., et al. (2020). "The Lancaster Sensorimotor Norms: Multidimensional measures of Perceptual and Action Strength for 40,000 English words." Behavior Research Methods52(3): 1271-1291. |
| Semantic | Specific Semantic Features | Head_Lanc | Lancaster | Head | Lynott, D., et al. (2020). "The Lancaster Sensorimotor Norms: Multidimensional measures of Perceptual and Action Strength for 40,000 English words." Behavior Research Methods52(3): 1271-1291. |
| Semantic | Specific Semantic Features | Torso_Lanc | Lancaster | Torso | Lynott, D., et al. (2020). "The Lancaster Sensorimotor Norms: Multidimensional measures of Perceptual and Action Strength for 40,000 English words." Behavior Research Methods52(3): 1271-1291. |
| Semantic | Specific Semantic Features | Mouth_Throat_Lanc | Lancaster | Mouth/throat | Lynott, D., et al. (2020). "The Lancaster Sensorimotor Norms: Multidimensional measures of Perceptual and Action Strength for 40,000 English words." Behavior Research Methods52(3): 1271-1291. |

| Semantic | Specific Semantic Features | Hand_Arm_L anc | Lancaster | Hand/arm | Lynott, D., et al. (2020). "The Lancaster Sensorimotor Norms: Multidimensional measures of Perceptual and Action Strength for 40,000 English words." Behavior Research Methods52(3): 1271-1291. |
|---|---|---|---|---|---|
| Semantic | Specific Semantic Features | Foot_Leg_La nc | Lancaster | Foot/leg | Lynott, D., et al. (2020). "The Lancaster Sensorimotor Norms: Multidimensional measures of Perceptual and Action Strength for 40,000 English words." Behavior Research Methods52(3): 1271-1291. |
| Semantic | Specific Semantic Features | Mink_Percept ual_Lanc | Lancaster | Minkowski3.perceptual | Lynott, D., et al. (2020). "The Lancaster Sensorimotor Norms: Multidimensional measures of Perceptual and Action Strength for 40,000 English words." Behavior Research Methods52(3): 1271-1291. |
| Semantic | Specific Semantic Features | Mink_Action _Lanc | Lancaster | Minkowski3.action | Lynott, D., et al. (2020). "The Lancaster Sensorimotor Norms: Multidimensional measures of Perceptual and Action Strength for 40,000 English words." Behavior Research Methods52(3): 1271-1291. |
| Semantic | Affect | Valence_NR C | NRC | | Mohammad, S. and P. Turney (2010). Emotions evoked by common words and phrases: Using mechanical turk to create an emotion lexicon. Proceedings of the NAACL HLT 2010 workshop on computational approaches to analysis and generation of emotion in text. Mohammad, S. M. and P. D. Turney (2013). "Crowdsourcing a word–emotion association lexicon." Computational intelligence 29(3): 436-465. |
| Semantic | Affect | Arousal_NRC | NRC | | Mohammad, S. and P. Turney (2010). Emotions evoked by common words and phrases: Using mechanical turk to create an emotion lexicon. Proceedings of the NAACL HLT 2010 workshop on computational approaches to analysis and generation of emotion in text. Mohammad, S. M. and P. D. Turney (2013). "Crowdsourcing a word–emotion association lexicon." Computational intelligence 29(3): 436-465. |
| Semantic | Affect | Dominance_ NRC | NRC | | Mohammad, S. and P. Turney (2010). Emotions evoked by common words and phrases: Using mechanical turk to create an emotion lexicon. Proceedings of the NAACL HLT 2010 workshop on computational approaches to analysis and generation of emotion in text. Mohammad, S. M. and P. D. Turney (2013). "Crowdsourcing a word–emotion association lexicon." Computational intelligence |
| | | | | | 29(3): 436-465. |
| Semantic | Semantic Neighborhood | Sem_Diversity | Hoffman13 | SemD | Hoffman, P., et al. (2013). "Semantic diversity: A measure of semantic ambiguity based on variability in the contextual usage of words." Behavior Research Methods45(3): 718-730. |
| Semantic | Semantic Neighborhood | Sem_N | HiDEx | NCOUNT | Shaoul, C. and C. Westbury (2006). "Word frequency effects in high-dimensional co-occurrence models: A new approach." Behavior Research Methods38(2): 190-195. Shaoul, C. and C. Westbury (2010). "Exploring lexical co-occurrence space using HiDEx." Behavior Research Methods 42(2): 393-413. |
| Semantic | Semantic Neighborhood | Sem_N_D | HiDEx | ARC | Shaoul, C. and C. Westbury (2006). "Word frequency effects in high-dimensional co-occurrence models: A new approach." Behavior Research Methods38(2): 190-195. Shaoul, C. and C. Westbury (2010). "Exploring lexical co-occurrence space using HiDEx." Behavior Research Methods 42(2): 393-413. |
| Semantic | Semantic Neighborhood | Sem_N_D_Taxonomic_N3 | Reilly17 | N3 | Reilly, M. and R. H. Desai (2017). "Effects of semantic neighborhood density in abstract and concrete words." Cognition169: 46-53. Roller, S. and K. Erk (2016). "Relations such as hypernymy: Identifying and exploiting hearst patterns in distributional vectors for lexical entailment." arXiv preprint arXiv:1605.05433. |
| Semantic | Semantic Neighborhood | Sem_N_D_Taxonomic_N10 | Reilly17 | N10 | Reilly, M. and R. H. Desai (2017). "Effects of semantic neighborhood density in abstract and concrete words." Cognition169: 46-53. Roller, S. and K. Erk (2016). "Relations such as hypernymy: Identifying and exploiting hearst patterns in distributional vectors for lexical entailment." arXiv preprint arXiv:1605.05433. |
| Semantic | Semantic Neighborhood | Sem_N_D_Taxonomic_N25 | Reilly17 | N25 | Reilly, M. and R. H. Desai (2017). "Effects of semantic neighborhood density in abstract and concrete words." Cognition169: 46-53. Roller, S. and K. Erk (2016). "Relations such as hypernymy: Identifying and exploiting hearst patterns in distributional vectors for lexical entailment." arXiv preprint arXiv:1605.05433. |

| Semantic | Semantic Neighborhood | Sem_N_D_Taxonomic_N50 | Reilly17 | N50 | Reilly, M. and R. H. Desai (2017). "Effects of semantic neighborhood density in abstract and concrete words." Cognition169: 46-53. Roller, S. and K. Erk (2016). "Relations such as hypernymy: Identifying and exploiting hearst patterns in distributional vectors for lexical entailment." arXiv preprint arXiv:1605.05433. |
|---|---|---|---|---|---|
| Semantic | Semantic Neighborhood | Assoc_Freq_Token | ELP | Assoc_Freq_R1 | De Deyne, S., et al. (2019). "The "Small World of Words" English word association norms for over 12,000 cue words." Behavior Research Methods51(3): 987-1006. |
| Semantic | Semantic Neighborhood | Assoc_Freq_Type | ELP | Assoc_Types_R1 | De Deyne, S., et al. (2019). "The "Small World of Words" English word association norms for over 12,000 cue words." Behavior Research Methods51(3): 987-1006. |
| Semantic | Semantic Neighborhood | Assoc_Freq_Token123 | ELP | Assoc_Freq_R123 | De Deyne, S., et al. (2019). "The "Small World of Words" English word association norms for over 12,000 cue words." Behavior Research Methods51(3): 987-1006. |
| Semantic | Semantic Neighborhood | Assoc_Freq_Type123 | ELP | Assoc_Types_R123 | De Deyne, S., et al. (2019). "The "Small World of Words" English word association norms for over 12,000 cue words." Behavior Research Methods51(3): 987-1006. |
| Orth-Phon | Phonographic Neighborhood | Phonographic_N | ELP | OG_N | Balota, D. A., et al. (2007). "The English lexicon project." Behavior Research Methods39(3): 445-459. |
| Orth-Phon | Consistency | Consistency_Token_FF_O | Chee20 | Token consistency: Feedforward consistency | Chee, Q. W., et al. (2020). "Consistency norms for 37,677 english words." Behavior Research Methods. |
| Orth-Phon | Consistency | Consistency_Token_FF_N | Chee20 | Token consistency: Feedforward consistency | Chee, Q. W., et al. (2020). "Consistency norms for 37,677 english words." Behavior Research Methods. |
| Orth-Phon | Consistency | Consistency_Token_FF_C | Chee20 | Token consistency: Feedforward consistency | Chee, Q. W., et al. (2020). "Consistency norms for 37,677 english words." Behavior Research Methods. |

| Orth-Phon | Consistency | Consistency_Token_FF_ON | Chee20 | Token consistency: Feedforward consistency | Chee, Q. W., et al. (2020). "Consistency norms for 37,677 english words." Behavior Research Methods. |
|---|---|---|---|---|---|
| Orth-Phon | Consistency | Consistency_Token_FF_R | Chee20 | Token consistency: Feedforward consistency | Chee, Q. W., et al. (2020). "Consistency norms for 37,677 english words." Behavior Research Methods. |
| Orth-Phon | Consistency | Consistency_Token_FF | Chee20 | Token consistency: Feedforward consistency | Chee, Q. W., et al. (2020). "Consistency norms for 37,677 english words." Behavior Research Methods. |
| Orth-Phon | Consistency | Consistency_Token_FB_O | Chee20 | Token consistency: Feedback consistency | Chee, Q. W., et al. (2020). "Consistency norms for 37,677 english words." Behavior Research Methods. |
| Orth-Phon | Consistency | Consistency_Token_FB_N | Chee20 | Token consistency: Feedback consistency | Chee, Q. W., et al. (2020). "Consistency norms for 37,677 english words." Behavior Research Methods. |
| Orth-Phon | Consistency | Consistency_Token_FB_C | Chee20 | Token consistency: Feedback consistency | Chee, Q. W., et al. (2020). "Consistency norms for 37,677 english words." Behavior Research Methods. |
| Orth-Phon | Consistency | Consistency_Token_FB_ON | Chee20 | Token consistency: Feedback consistency | Chee, Q. W., et al. (2020). "Consistency norms for 37,677 english words." Behavior Research Methods. |
| Orth-Phon | Consistency | Consistency_Token_FB_R | Chee20 | Token consistency: Feedback consistency | Chee, Q. W., et al. (2020). "Consistency norms for 37,677 english words." Behavior Research Methods. |
| Orth-Phon | Consistency | Consistency_Token_FB | Chee20 | Token consistency: Feedback consistency | Chee, Q. W., et al. (2020). "Consistency norms for 37,677 english words." Behavior Research Methods. |
| Orth-Phon | Consistency | Consistency_Type_FF_O | Chee20 | Type consistency: Feedforward consistency | Chee, Q. W., et al. (2020). "Consistency norms for 37,677 english words." Behavior Research Methods. |
| Orth-Phon | Consistency | Consistency_Type_FF_N | Chee20 | Type consistency: Feedforward consistency | Chee, Q. W., et al. (2020). "Consistency norms for 37,677 english words." Behavior Research Methods. |
| Orth-Phon | Consistency | Consistency_Type_FF_C | Chee20 | Type consistency: Feedforward consistency | Chee, Q. W., et al. (2020). "Consistency norms for 37,677 english words." Behavior Research Methods. |

| Orth-Phon | Consistency | Consistency_Type_FF_ON | Chee20 | Type consistency: Feedforward consistency | Chee, Q. W., et al. (2020). "Consistency norms for 37,677 english words." Behavior Research Methods. |
|---|---|---|---|---|---|
| Orth-Phon | Consistency | Consistency_Type_FF_R | Chee20 | Type consistency: Feedforward consistency | Chee, Q. W., et al. (2020). "Consistency norms for 37,677 english words." Behavior Research Methods. |
| Orth-Phon | Consistency | Consistency_Type_FF | Chee20 | Type consistency: Feedforward consistency | Chee, Q. W., et al. (2020). "Consistency norms for 37,677 english words." Behavior Research Methods. |
| Orth-Phon | Consistency | Consistency_Type_FB_O | Chee20 | Type consistency: Feedback consistency | Chee, Q. W., et al. (2020). "Consistency norms for 37,677 english words." Behavior Research Methods. |
| Orth-Phon | Consistency | Consistency_Type_FB_N | Chee20 | Type consistency: Feedback consistency | Chee, Q. W., et al. (2020). "Consistency norms for 37,677 english words." Behavior Research Methods. |
| Orth-Phon | Consistency | Consistency_Type_FB_C | Chee20 | Type consistency: Feedback consistency | Chee, Q. W., et al. (2020). "Consistency norms for 37,677 english words." Behavior Research Methods. |
| Orth-Phon | Consistency | Consistency_Type_FB_ON | Chee20 | Type consistency: Feedback consistency | Chee, Q. W., et al. (2020). "Consistency norms for 37,677 english words." Behavior Research Methods. |
| Orth-Phon | Consistency | Consistency_Type_FB_R | Chee20 | Type consistency: Feedback consistency | Chee, Q. W., et al. (2020). "Consistency norms for 37,677 english words." Behavior Research Methods. |
| Orth-Phon | Consistency | Consistency_Type_FB | Chee20 | Type consistency: Feedback consistency | Chee, Q. W., et al. (2020). "Consistency norms for 37,677 english words." Behavior Research Methods. |
| Morphology | Morphological Length | NMorph | ELP | NMorph | Sánchez-Gutiérrez, C. H., et al. (2018). "MorphoLex: A derivational morphological database for 70,000 English words." Behavior Research Methods50(4): 1568-1580. |
| Morphology | Frequency | ROOT1_Freq_HAL | MorphoLex | ROOT1_Freq_HAL | Sánchez-Gutiérrez, C. H., et al. (2018). "MorphoLex: A derivational morphological database for 70,000 English words." Behavior Research Methods50(4): 1568-1580. |
| Morphology | Family size | ROOT1_Fam Size | MorphoLex | ROOT1_FamSize | Sánchez-Gutiérrez, C. H., et al. (2018). "MorphoLex: A derivational morphological database for 70,000 English words." Behavior Research Methods50(4): 1568-1580. |
| Morphology | Percent more frequent | ROOT1_PFMF | MorphoLex | ROOT1_PFMF | Sánchez-Gutiérrez, C. H., et al. (2018). "MorphoLex: A derivational morphological database for 70,000 English words." Behavior Research Methods50(4): 1568-1580. |

| Response Variables | Visual Lexical Decision | LexicalD_RT_V_ELP_z | ELP | I_Mean_RT | Balota, D. A., et al. (2007). "The English lexicon project." Behavior Research Methods39(3): 445-459. |
|---|---|---|---|---|---|
| Response Variables | Visual Lexical Decision | LexicalD_RT_V_ECP_z | ECP | rt_mean | Mandera, P., et al. (2019). "Recognition times for 62 thousand English words: Data from the English Crowdsourcing Project." Behavior Research Methods: 1-20. |
| Response Variables | Visual Lexical Decision | LexicalD_RT_V_BLP_z | BLP | rt | Keuleers, E., et al. (2012). "The British Lexicon Project: Lexical decision data for 28,730 monosyllabic and disyllabic English words." Behavior Research Methods44(1): 287-304. |
| Response Variables | Visual Lexical Decision | LexicalD_ACC_V_ELP | ELP | I_Mean_Accuracy | Balota, D. A., et al. (2007). "The English lexicon project." Behavior Research Methods39(3): 445-459. |
| Response Variables | Visual Lexical Decision | LexicalD_ACC_V_ECP | ECP | accuracy | Mandera, P., et al. (2019). "Recognition times for 62 thousand English words: Data from the English Crowdsourcing Project." Behavior Research Methods: 1-20. |
| Response Variables | Visual Lexical Decision | LexicalD_ACC_V_BLP | BLP | accuracy | Keuleers, E., et al. (2012). "The British Lexicon Project: Lexical decision data for 28,730 monosyllabic and disyllabic English words." Behavior Research Methods44(1): 287-304. |
| Response Variables | Auditory Lexical Decision | LexicalD_RT_A_MALD_z | MALD | RT | Tucker, B. V., et al. (2019). "The massive auditory lexical decision (MALD) database." Behavior Research Methods51(3): 1187-1204. |
| Response Variables | Auditory Lexical Decision | LexicalD_RT_A_AELP_z | AELP | LDT Reaction Time | Goh, W. D., et al. (2020). "The Auditory English Lexicon Project: A multi-talker, multi-region psycholinguistic database of 10,170 spoken words and nonwords." Behavior Research Methods: 1-30. |
| Response Variables | Auditory Lexical Decision | LexicalD_ACC_A_MALD | MALD | ACC | Tucker, B. V., et al. (2019). "The massive auditory lexical decision (MALD) database." Behavior Research Methods 51(3): 1187-1204. |
| Response Variables | Auditory Lexical Decision | LexicalD_ACC_A_AELP | AELP | LDT Accuracy | Goh, W. D., et al. (2020). "The Auditory English Lexicon Project: A multi-talker, multi-region psycholinguistic database of 10,170 spoken words and nonwords." Behavior Research Methods: 1-30. |
| Response Variables | Reading Aloud | Naming_RT_ELP_z | ELP | I_NMG_Mean_RT | Balota, D. A., et al. (2007). "The English lexicon project." Behavior Research Methods 39(3): 445-459. |
| Response Variables | Reading Aloud | Naming_ACC_ELP | ELP | I_NMG_Mean_Accuracy | Balota, D. A., et al. (2007). "The English lexicon project." Behavior Research Methods 39(3): 445-459. |
| Response Variables | Semantic Decision | SemanticD_RT_Calgary_z | Calgary | RTclean_mean | Pexman, P. M., et al. (2017). "The Calgary semantic decision project: concrete/abstract decision data for 10,000 English words." Behavior Research Methods 49(2): 407-417. |
| Response Variables | Semantic Decision | SemanticD_ACC_Calgary | Calgary | ACC | Pexman, P. M., et al. (2017). "The Calgary semantic decision project: concrete/abstract decision data for 10,000 English words." Behavior Research Methods 49(2): 407-417. |
| Response Variables | Recognition Memory | Recog_Memory | Khanna21 | hitsmfas | Cortese, M. J., Khanna, M. M., & Hacker, S. (2010). Recognition memory for 2,578 monosyllabic words. Memory, 18(6), 595-609; Cortese, M. J., McCarty, D. P., & Schock, J. (2015). A mega recognition memory study of 2897 disyllabic words. Quarterly Journal of Experimental Psychology, 68(8), 1489-1501; Khanna, M. M. and M. J. Cortese (2021). "How well imageability, concreteness, perceptual strength, and action strength predict recognition memory, lexical decision, and reading aloud performance." Memory: 1-15. |

**Table S2**: Factor loadings

| Variable | FA1 | FA2 | FA3 | FA4 | FA5 | FA6 | FA7 | FA8 |
|---|---|---|---|---|---|---|---|---|
| CD_Blog_LemmaSum | **0.958** | 0.008 | 0.004 | 0.000 | 0.003 | -0.019 | 0.022 | 0.013 |
| Freq_Blog_LemmaSum | **0.955** | 0.005 | 0.002 | -0.004 | 0.004 | -0.008 | 0.021 | 0.016 |
| CD_SUBTLEXUK_LemmaSum | **0.953** | -0.015 | 0.007 | 0.018 | -0.004 | -0.012 | 0.016 | -0.010 |
| CD_News_LemmaSum | **0.944** | 0.059 | 0.010 | -0.015 | 0.012 | -0.017 | 0.099 | -0.079 |
| Freq_SUBTLEXUK_Zipf_LemmaSum | **0.943** | -0.022 | -0.004 | 0.003 | -0.005 | 0.026 | -0.001 | 0.006 |
| Freq_SUBTLEXUK_LemmaSum | **0.943** | -0.022 | -0.004 | 0.003 | -0.005 | 0.026 | -0.001 | 0.006 |
| Freq_News_LemmaSum | **0.942** | 0.056 | 0.009 | -0.016 | 0.012 | -0.010 | 0.097 | -0.077 |
| CD_Twitter_LemmaSum | **0.920** | -0.035 | 0.007 | 0.013 | -0.027 | -0.003 | -0.053 | 0.061 |
| Freq_Twitter_LemmaSum | **0.919** | -0.037 | 0.007 | 0.013 | -0.027 | -0.002 | -0.054 | 0.062 |
| CD_SUBTLEXUS_LemmaSum | **0.904** | -0.071 | -0.012 | 0.016 | -0.011 | -0.001 | -0.074 | 0.069 |
| Freq_SUBTLEXUS_LemmaSum | **0.897** | -0.083 | -0.019 | 0.003 | -0.011 | 0.011 | -0.087 | 0.079 |
| Freq_SUBTLEXUS_Zipf_LemmaSum | **0.897** | -0.083 | -0.019 | 0.003 | -0.011 | 0.011 | -0.087 | 0.079 |
| Freq_HAL_LemmaSum | **0.897** | 0.021 | 0.005 | -0.043 | 0.035 | -0.067 | 0.097 | -0.057 |
| Freq_KF_LemmaSum | **0.861** | 0.005 | -0.010 | -0.018 | 0.037 | -0.086 | 0.079 | -0.030 |
| Cumfreq_TASA | **0.612** | -0.096 | -0.033 | -0.024 | -0.002 | 0.111 | 0.133 | 0.191 |
| Sem_N | **0.578** | 0.071 | -0.031 | -0.047 | 0.002 | -0.043 | 0.367 | 0.002 |
| Nsenses_WordNet | **0.491** | -0.184 | 0.103 | 0.022 | 0.023 | 0.062 | 0.084 | -0.093 |
| Sem_N_D | **0.463** | 0.151 | -0.042 | -0.090 | -0.016 | 0.023 | 0.357 | -0.115 |
| Sem_Diversity | **0.421** | 0.013 | 0.032 | 0.036 | 0.012 | -0.320 | 0.204 | -0.043 |
| AoA_Kuper | **-0.524** | 0.108 | 0.001 | -0.074 | 0.054 | **-0.421** | 0.178 | -0.156 |
| AoA_LWV | **-0.531** | 0.136 | -0.005 | -0.082 | 0.057 | **-0.420** | 0.190 | -0.157 |
| OLD20 | -0.059 | **0.884** | -0.086 | 0.001 | -0.092 | 0.001 | 0.020 | -0.040 |
| PLD20 | -0.067 | **0.877** | 0.070 | -0.092 | -0.075 | -0.013 | 0.034 | -0.029 |
| NLett | -0.005 | **0.813** | 0.032 | 0.025 | 0.167 | -0.055 | -0.092 | -0.040 |
| OUP | 0.028 | **0.806** | -0.015 | -0.001 | 0.123 | -0.001 | 0.089 | -0.055 |
| NPhon | -0.047 | **0.799** | 0.143 | -0.056 | 0.194 | -0.054 | -0.030 | -0.031 |
| PUP | -0.030 | **0.742** | 0.106 | -0.045 | 0.231 | -0.017 | 0.129 | -0.021 |
| NSyll | -0.058 | **0.729** | 0.025 | -0.092 | 0.229 | -0.090 | 0.013 | 0.021 |
| NMorph | 0.107 | **0.548** | 0.045 | 0.026 | 0.070 | -0.031 | -0.275 | -0.143 |
| Consistency_Token_FB_C | -0.030 | **0.407** | 0.158 | 0.146 | 0.294 | 0.013 | 0.001 | 0.083 |
| Consistency_Type_FB_C | -0.035 | **0.405** | 0.156 | 0.149 | 0.293 | -0.012 | 0.003 | 0.085 |
| BigramF_Avg_C_Log | 0.134 | **-0.625** | 0.091 | -0.067 | 0.325 | -0.138 | -0.127 | 0.059 |
| Phonographic_N | 0.034 | **-0.683** | 0.198 | 0.059 | 0.057 | 0.063 | 0.021 | -0.023 |
| OLD20F | 0.016 | **-0.707** | 0.017 | -0.133 | 0.001 | -0.057 | 0.123 | 0.108 |
| PLD20F | 0.093 | **-0.737** | -0.081 | -0.092 | 0.015 | -0.080 | 0.097 | 0.053 |
| Orth_N | 0.065 | **-0.739** | 0.165 | -0.058 | 0.048 | 0.067 | 0.013 | -0.008 |
| Orth_Spread | 0.050 | **-0.742** | 0.144 | 0.011 | 0.059 | 0.041 | -0.027 | 0.014 |
| Phon_N | 0.076 | **-0.753** | -0.097 | 0.031 | 0.034 | 0.048 | 0.036 | -0.019 |
| Phon_Spread | 0.061 | **-0.753** | -0.038 | 0.137 | 0.084 | 0.054 | -0.045 | 0.001 |
| UnigramF_Avg_C_Log | 0.050 | **-0.855** | -0.028 | -0.065 | 0.013 | 0.005 | 0.010 | 0.028 |
| Consistency_Token_FB | -0.019 | 0.080 | **0.968** | 0.037 | 0.029 | -0.018 | 0.016 | 0.032 |
| Consistency_Type_FB | -0.030 | 0.077 | **0.966** | 0.049 | 0.030 | -0.011 | 0.014 | 0.035 |
| Consistency_Token_FB_ON | -0.015 | -0.045 | **0.811** | -0.021 | -0.150 | -0.043 | 0.062 | 0.006 |
| Consistency_Type_FB_ON | -0.027 | -0.052 | **0.811** | -0.013 | -0.146 | -0.037 | 0.059 | 0.009 |
| Consistency_Token_FB_N | 0.020 | -0.098 | **0.810** | -0.082 | 0.036 | 0.043 | -0.026 | -0.065 |
| Consistency_Type_FB_N | 0.019 | -0.098 | **0.810** | -0.070 | 0.034 | 0.046 | -0.027 | -0.066 |
| Consistency_Token_FB_R | 0.046 | 0.091 | **0.768** | 0.046 | 0.020 | 0.077 | -0.060 | -0.016 |
| Consistency_Type_FB_R | 0.036 | 0.092 | **0.765** | 0.055 | 0.020 | 0.084 | -0.061 | -0.012 |
| Consistency_Token_FF | -0.006 | -0.141 | 0.080 | **0.948** | -0.035 | -0.007 | 0.016 | -0.022 |
| Consistency_Type_FF | -0.017 | -0.145 | 0.089 | **0.945** | -0.035 | -0.001 | 0.012 | -0.019 |
| Consistency_Token_FF_ON | 0.022 | 0.116 | -0.088 | **0.889** | -0.010 | -0.025 | 0.021 | 0.012 |
| Consistency_Type_FF_ON | 0.006 | 0.105 | -0.080 | **0.889** | -0.005 | -0.016 | 0.016 | 0.015 |
| Consistency_Type_FF_N | -0.008 | 0.116 | -0.022 | **0.846** | 0.141 | -0.004 | -0.001 | -0.003 |
| Consistency_Token_FF_N | -0.006 | 0.120 | -0.031 | **0.846** | 0.144 | -0.009 | 0.000 | -0.002 |
| Consistency_Token_FF_R | 0.027 | -0.340 | 0.138 | **0.485** | -0.122 | 0.016 | -0.003 | -0.083 |
| Consistency_Type_FF_R | 0.021 | -0.341 | 0.145 | **0.481** | -0.125 | 0.018 | -0.006 | -0.080 |
| UniphonP_St_C | 0.056 | -0.107 | -0.006 | 0.022 | **0.820** | 0.055 | -0.111 | -0.048 |
| UniphonP_Un_C | 0.054 | -0.018 | -0.019 | 0.036 | **0.819** | 0.035 | -0.099 | -0.044 |
| UniphonP_Un | -0.077 | -0.065 | -0.106 | 0.078 | **0.805** | 0.096 | 0.062 | -0.012 |
| UniphonP_St | -0.081 | -0.117 | -0.078 | 0.026 | **0.799** | 0.124 | 0.055 | -0.025 |
| BiphonP_St | -0.030 | 0.126 | 0.017 | -0.035 | **0.757** | -0.081 | 0.038 | 0.013 |
| BiphonP_Un | -0.044 | 0.114 | -0.025 | 0.088 | **0.734** | -0.066 | 0.042 | 0.040 |
| TriphonP_Un | -0.026 | 0.145 | 0.028 | 0.028 | **0.720** | -0.051 | 0.106 | 0.031 |
| TriphonP_St | -0.012 | 0.145 | 0.036 | -0.023 | **0.711** | -0.060 | 0.099 | 0.017 |
| TrigramF_Avg_U_Log | 0.163 | 0.120 | 0.182 | -0.115 | **0.461** | -0.190 | -0.137 | 0.046 |
| TrigramF_Avg_C_Log | 0.252 | -0.236 | 0.137 | -0.042 | **0.453** | -0.195 | -0.132 | 0.099 |
| BigramF_Avg_U_Log | 0.075 | 0.031 | 0.137 | 0.147 | **0.432** | -0.162 | -0.099 | 0.037 |
| Conc_Brys | -0.092 | -0.116 | -0.005 | -0.020 | -0.018 | **0.802** | -0.011 | -0.040 |
| Visual_Lanc | 0.017 | 0.088 | 0.009 | -0.051 | 0.008 | **0.759** | 0.094 | -0.048 |
| Haptic_Lanc | -0.021 | -0.047 | 0.070 | 0.014 | 0.051 | **0.749** | -0.003 | 0.041 |
| Mink_Perceptual_Lanc | -0.013 | 0.128 | -0.015 | 0.008 | -0.017 | **0.641** | -0.098 | 0.357 |
| Hand_Arm_Lanc | 0.111 | 0.080 | 0.105 | 0.032 | 0.025 | **0.585** | 0.030 | 0.043 |
| Interoceptive_Lanc | 0.028 | 0.105 | 0.017 | 0.062 | -0.055 | **-0.426** | -0.158 | **0.446** |
| Sem_N_D_Taxonomic_N10 | 0.033 | -0.012 | 0.005 | 0.016 | 0.006 | 0.005 | **0.923** | 0.056 |
| Sem_N_D_Taxonomic_N25 | 0.031 | -0.011 | 0.003 | 0.015 | 0.007 | 0.007 | **0.922** | 0.063 |
| Sem_N_D_Taxonomic_N50 | 0.029 | -0.010 | 0.002 | 0.014 | 0.003 | 0.010 | **0.920** | 0.068 |
| Sem_N_D_Taxonomic_N3 | 0.036 | -0.016 | 0.009 | 0.018 | 0.008 | 0.003 | **0.920** | 0.048 |
| Assoc_Freq_Token123 | 0.005 | -0.090 | -0.002 | -0.029 | 0.014 | 0.036 | 0.129 | **0.800** |
| Assoc_Freq_Type | 0.074 | -0.092 | 0.001 | -0.024 | 0.008 | 0.018 | 0.121 | **0.784** |
| Assoc_Freq_Type123 | 0.059 | -0.088 | -0.004 | -0.020 | 0.007 | 0.017 | 0.154 | **0.779** |
| Assoc_Freq_Token | 0.018 | -0.091 | -0.003 | -0.034 | 0.014 | 0.033 | 0.097 | **0.779** |
| Mink_Action_Lanc | 0.154 | 0.242 | 0.019 | 0.062 | -0.077 | 0.181 | -0.133 | **0.482** |
| Mouth_Throat_Lanc | 0.039 | 0.175 | -0.064 | 0.035 | -0.040 | -0.299 | -0.165 | **0.475** |
| Freqtraj_TASA | 0.118 | 0.358 | -0.045 | -0.111 | 0.059 | -0.293 | 0.278 | -0.196 |
| Fam_Brys | 0.228 | 0.214 | -0.017 | 0.032 | -0.089 | 0.198 | 0.039 | 0.012 |
| Prevalence_Brys | 0.338 | 0.306 | 0.005 | 0.061 | -0.098 | 0.162 | 0.071 | -0.004 |
| UnigramF_Avg_U_Log | 0.015 | 0.047 | -0.054 | -0.190 | 0.269 | -0.099 | 0.070 | -0.028 |
| Nmeanings_Websters | 0.348 | -0.135 | 0.028 | 0.002 | 0.026 | 0.043 | 0.156 | -0.068 |
| Head_Lanc | 0.136 | 0.310 | -0.041 | -0.020 | -0.091 | -0.221 | -0.034 | 0.364 |
| Auditory_Lanc | 0.106 | 0.185 | -0.058 | 0.045 | -0.024 | -0.325 | -0.071 | 0.259 |
| Gustatory_Lanc | -0.056 | 0.052 | -0.025 | -0.017 | 0.016 | 0.197 | -0.130 | 0.341 |
| Olfactory_Lanc | -0.083 | 0.069 | -0.027 | -0.052 | 0.009 | 0.317 | -0.093 | 0.340 |
| Torso_Lanc | 0.053 | 0.132 | 0.029 | 0.046 | -0.028 | 0.158 | -0.044 | 0.316 |
| Foot_Leg_Lanc | 0.100 | 0.093 | 0.049 | 0.050 | -0.030 | 0.225 | 0.038 | 0.129 |
| Valence_NRC | 0.291 | 0.265 | -0.024 | -0.042 | -0.036 | 0.069 | 0.073 | -0.031 |
| Arousal_NRC | -0.066 | 0.106 | 0.019 | 0.074 | -0.034 | -0.333 | -0.063 | 0.228 |
| Dominance_NRC | 0.192 | 0.330 | -0.007 | -0.036 | -0.007 | -0.276 | 0.134 | -0.014 |
| Consistency_Token_FF_C | -0.057 | -0.037 | 0.153 | 0.113 | 0.079 | -0.018 | 0.091 | -0.013 |
| Consistency_Type_FF_C | -0.060 | -0.035 | 0.155 | 0.114 | 0.077 | -0.017 | 0.092 | -0.010 |
| Consistency_Token_FB_O | -0.097 | -0.246 | 0.294 | -0.006 | -0.209 | -0.153 | 0.091 | 0.097 |
| Consistency_Type_FB_O | -0.098 | -0.246 | 0.295 | -0.005 | -0.209 | -0.151 | 0.090 | 0.097 |
| Consistency_Type_FF_O | -0.044 | -0.302 | 0.148 | 0.256 | -0.170 | 0.010 | -0.030 | 0.034 |
| Consistency_Token_FF_O | -0.043 | -0.302 | 0.148 | 0.257 | -0.168 | 0.011 | -0.029 | 0.034 |
| ROOT1_PFMF | -0.224 | 0.230 | -0.006 | 0.013 | -0.023 | -0.072 | 0.022 | 0.019 |
| ROOT1_FamSize | 0.216 | 0.147 | 0.083 | -0.020 | -0.066 | 0.087 | 0.039 | 0.026 |
| ROOT1_Freq_HAL | 0.378 | 0.122 | 0.030 | -0.029 | -0.074 | -0.079 | 0.003 | -0.079 |

*Notes.* Loadings with absolute values greater than 0.4 are shown in bold.

**Table S3: The top five words with the highest positive and negative factor scores**

|          | F1: Word Frequency | F2: Word Length | F3: Feedback Consistency | F4: Feedforward Consistency |
|----------|--------------------|-----------------|--------------------------|-----------------------------|
| **Positive** | are/be | bibliographies/bibliography | man/man | squealed/squeal |
|          | was/be | manufacturers/manufacturer | hands/hand | liars/liar |
|          | being/be | relationships/relationship | bagging/bag | liar/liar |
|          | were/be | grandchildren/grandchild | stands/stand | sleeps/sleep |
|          | been/be | enthusiastic/enthusiastic | banked/bank | swum/swim |
| **Negative** | homogeneous/homogeneous | may/may | twos/two | cough/cough |
|          | heredity/heredity | mane/mane | knew/know | one/one |
|          | bluish/bluish | hay/hay | one/one | hour/hour |
|          | weariness/weariness | cane/cane | rhyme/rhyme | coughs/cough |
|          | linoleum/linoleum | say/say | suite/suite | was/be |
|          | **F5: Graphotactic/phonotactic Probability** | **F6: Concreteness** | **F7: Taxonomic Semantic Neighborhood** | **F8: Associative Semantic Neighborhood** |
| **Positive** | contention/contention | puppy/puppy | occur/occur | food/food |
|          | continent/continent | soaps/soap | form/form | money/money |
|          | contentions/contention | soap/soap | feature/feature | monies/money |
|          | continents/continent | sofas/sofa | portrayed/portray | water/water |
|          | scenting/scent | pumpkin/pumpkin | compare/compare | love/love |
| **Negative** | ooze/ooze | witting/wit | coughing/cough | modes/mode |
|          | joy/joy | divining/divine | kidding/kid | enzymes/enzyme |
|          | vow/vow | theoretical/theoretical | sneezing/sneeze | wholesale/wholesale |
|          | oak/oak | was/be | laughs/laugh | enzyme/enzyme |
|          | egg/egg | eminent/eminent | licking/lick | centuries/century |

*Notes*. The raw word form is shown before the slash, and the corresponding lemma is shown after the slash.

## Table S4: Best predictors

| Outcome variable (Z) | Best single predictor (X) | Spearman $r_{XZ}$ | Factor (Y) | Factor rank | Spearman $r_{YZ}$ | Steiger's $T$ | $N$ | $p$ value | $p$ value (FDR) |
|---|---|---|---|---|---|---|---|---|---|
| LexicalD_RT_V_ELP_z | Cumfreq_TASA | 0.49 | FA1 | 2 | 0.49 | 0.52 | 11617 | 0.603 | 0.629 |
| LexicalD_RT_V_ELP_z | NLett | 0.52 | FA2 | 6 | 0.44 | 22.47 | 11617 | 0.000 | 0.000 |
| LexicalD_RT_V_ELP_z | Consistency_Token_FB_ON | 0.04 | FA3 | 8 | 0.00 | 6.35 | 11617 | 0.000 | 0.000 |
| LexicalD_RT_V_ELP_z | Consistency_Token_FF_R | 0.20 | FA4 | 6 | 0.05 | 16.58 | 11617 | 0.000 | 0.000 |
| LexicalD_RT_V_ELP_z | BiphonP_Un | 0.23 | FA5 | 3 | 0.21 | 2.07 | 11617 | 0.039 | 0.055 |
| LexicalD_RT_V_ELP_z | AoA_LWV | 0.47 | FA6 | 3 | 0.24 | 29.28 | 11617 | 0.000 | 0.000 |
| LexicalD_RT_V_ELP_z | Sem_N_D_Taxonomic_N50 | 0.16 | FA7 | 1 | 0.17 | -2.09 | 11617 | 0.037 | 0.053 |
| LexicalD_RT_V_ELP_z | Assoc_Freq_Token123 | 0.43 | FA8 | 5 | 0.29 | 17.36 | 11617 | 0.000 | 0.000 |
| LexicalD_RT_V_ECP_z | Sem_N_D | 0.52 | FA1 | 4 | 0.47 | 5.73 | 7449 | 0.000 | 0.000 |
| LexicalD_RT_V_ECP_z | NLett | 0.44 | FA2 | 11 | 0.20 | 43.22 | 7449 | 0.000 | 0.000 |
| LexicalD_RT_V_ECP_z | Consistency_Type_FB | 0.09 | FA3 | 1 | 0.09 | -0.87 | 7449 | 0.383 | 0.429 |
| LexicalD_RT_V_ECP_z | Consistency_Token_FF_R | 0.07 | FA4 | 7 | 0.01 | 5.19 | 7449 | 0.000 | 0.000 |
| LexicalD_RT_V_ECP_z | TrigramF_Avg_U_Log | 0.22 | FA5 | 1 | 0.24 | -1.57 | 7449 | 0.117 | 0.155 |
| LexicalD_RT_V_ECP_z | AoA_Kuper | 0.38 | FA6 | 4 | 0.21 | 15.23 | 7449 | 0.000 | 0.000 |
| LexicalD_RT_V_ECP_z | Sem_N_D_Taxonomic_N50 | 0.34 | FA7 | 1 | 0.34 | -0.30 | 7449 | 0.762 | 0.788 |
| LexicalD_RT_V_ECP_z | Assoc_Freq_Token123 | 0.57 | FA8 | 5 | 0.43 | 18.19 | 7449 | 0.000 | 0.000 |
| LexicalD_RT_V_BLP_z | Sem_N_D | 0.57 | FA1 | 3 | 0.51 | 8.49 | 9608 | 0.000 | 0.000 |
| LexicalD_RT_V_BLP_z | NMorph | 0.32 | FA2 | 12 | 0.09 | 20.96 | 9608 | 0.000 | 0.000 |
| LexicalD_RT_V_BLP_z | Consistency_Type_FB_N | 0.05 | FA3 | 1 | 0.05 | -0.86 | 9608 | 0.392 | 0.435 |
| LexicalD_RT_V_BLP_z | Consistency_Type_FF_N | 0.04 | FA4 | 5 | 0.03 | 2.01 | 9608 | 0.045 | 0.063 |
| LexicalD_RT_V_BLP_z | TrigramF_Avg_U_Log | 0.12 | FA5 | 3 | 0.09 | 3.09 | 9608 | 0.002 | 0.003 |
| LexicalD_RT_V_BLP_z | AoA_LWV | 0.37 | FA6 | 4 | 0.12 | 25.08 | 9608 | 0.000 | 0.000 |
| LexicalD_RT_V_BLP_z | Sem_N_D_Taxonomic_N50 | 0.24 | FA7 | 1 | 0.29 | -10.81 | 9608 | 0.000 | 0.000 |
| LexicalD_RT_V_BLP_z | Assoc_Freq_Type123 | 0.38 | FA8 | 5 | 0.30 | 9.50 | 9608 | 0.000 | 0.000 |
| LexicalD_RT_A_MALD_z | AoA_LWV | 0.21 | FA1 | 9 | 0.15 | 6.90 | 10720 | 0.000 | 0.000 |
| LexicalD_RT_A_MALD_z | NPhon | 0.18 | FA2 | 8 | 0.13 | 12.60 | 10720 | 0.000 | 0.000 |
| LexicalD_RT_A_MALD_z | Consistency_Token_FB_N | 0.02 | FA3 | 3 | 0.02 | 0.64 | 10720 | 0.525 | 0.558 |
| LexicalD_RT_A_MALD_z | Consistency_Token_FF | 0.06 | FA4 | 3 | 0.05 | 2.94 | 10720 | 0.003 | 0.005 |
| LexicalD_RT_A_MALD_z | BiphonP_St | 0.10 | FA5 | 4 | 0.08 | 3.15 | 10720 | 0.002 | 0.003 |
| LexicalD_RT_A_MALD_z | AoA_LWV | 0.21 | FA6 | 4 | 0.10 | 11.21 | 10720 | 0.000 | 0.000 |
| LexicalD_RT_A_MALD_z | Sem_N_D_Taxonomic_N50 | 0.04 | FA7 | 5 | 0.01 | 5.52 | 10720 | 0.000 | 0.000 |
| LexicalD_RT_A_MALD_z | Assoc_Freq_Token123 | 0.21 | FA8 | 3 | 0.20 | 1.21 | 10720 | 0.226 | 0.269 |
| LexicalD_RT_A_AELP_z | Freq_SUBTLEXUS_Zipf_LemmaSum | 0.32 | FA1 | 8 | 0.30 | 3.03 | 4495 | 0.002 | 0.004 |
| LexicalD_RT_A_AELP_z | NPhon | 0.26 | FA2 | 8 | 0.18 | 14.44 | 4495 | 0.000 | 0.000 |
| LexicalD_RT_A_AELP_z | Consistency_Type_FB_ON | 0.06 | FA3 | 1 | 0.06 | -0.02 | 4495 | 0.986 | 0.986 |
| LexicalD_RT_A_AELP_z | Consistency_Token_FF_N | 0.13 | FA4 | 3 | 0.12 | 1.37 | 4495 | 0.170 | 0.214 |
| LexicalD_RT_A_AELP_z | TrigramF_Avg_U_Log | 0.11 | FA5 | 8 | 0.07 | 3.14 | 4495 | 0.002 | 0.003 |
| LexicalD_RT_A_AELP_z | AoA_LWV | 0.29 | FA6 | 4 | 0.15 | 10.69 | 4495 | 0.000 | 0.000 |
| LexicalD_RT_A_AELP_z | Sem_N_D_Taxonomic_N50 | 0.14 | FA7 | 5 | 0.10 | 5.58 | 4495 | 0.000 | 0.000 |
| LexicalD_RT_A_AELP_z | Assoc_Freq_Token123 | 0.42 | FA8 | 5 | 0.32 | 10.61 | 4495 | 0.000 | 0.000 |
| Naming_RT_ELP_z | AoA_LWV | 0.43 | FA1 | 7 | 0.39 | 4.96 | 11617 | 0.000 | 0.000 |
| Naming_RT_ELP_z | NLett | 0.54 | FA2 | 5 | 0.47 | 19.34 | 11617 | 0.000 | 0.000 |
| Naming_RT_ELP_z | Consistency_Type_FB_N | 0.08 | FA3 | 5 | 0.07 | 2.38 | 11617 | 0.018 | 0.026 |
| Naming_RT_ELP_z | Consistency_Token_FF_R | 0.23 | FA4 | 5 | 0.10 | 14.31 | 11617 | 0.000 | 0.000 |
| Naming_RT_ELP_z | BiphonP_St | 0.28 | FA5 | 2 | 0.27 | 1.38 | 11617 | 0.169 | 0.214 |
| Naming_RT_ELP_z | AoA_LWV | 0.43 | FA6 | 3 | 0.21 | 25.96 | 11617 | 0.000 | 0.000 |
| Naming_RT_ELP_z | Sem_N_D_Taxonomic_N50 | 0.09 | FA7 | 5 | 0.09 | 0.66 | 11617 | 0.508 | 0.544 |
| Naming_RT_ELP_z | Assoc_Freq_Token123 | 0.33 | FA8 | 5 | 0.22 | 12.83 | 11617 | 0.000 | 0.000 |
| SemanticD_RT_Calgary_z | AoA_Kuper | 0.31 | FA1 | 14 | 0.10 | 11.66 | 2426 | 0.000 | 0.000 |
| SemanticD_RT_Calgary_z | NMorph | 0.19 | FA2 | 13 | 0.02 | 8.12 | 2426 | 0.000 | 0.000 |
| SemanticD_RT_Calgary_z | Consistency_Token_FB | 0.14 | FA3 | 2 | 0.13 | 0.13 | 2426 | 0.900 | 0.915 |
| SemanticD_RT_Calgary_z | Consistency_Token_FF_N | 0.04 | FA4 | 7 | 0.01 | 2.44 | 2426 | 0.015 | 0.023 |
| SemanticD_RT_Calgary_z | TrigramF_Avg_U_Log | 0.21 | FA5 | 3 | 0.19 | 1.28 | 2426 | 0.202 | 0.247 |
| SemanticD_RT_Calgary_z | Conc_Brys | 0.39 | FA6 | 2 | 0.38 | 1.34 | 2426 | 0.182 | 0.227 |
| SemanticD_RT_Calgary_z | Sem_N_D_Taxonomic_N50 | 0.11 | FA7 | 5 | 0.07 | 4.43 | 2426 | 0.000 | 0.000 |
| SemanticD_RT_Calgary_z | Assoc_Freq_Token123 | 0.34 | FA8 | 5 | 0.24 | 5.98 | 2426 | 0.000 | 0.000 |
| Recog_Memory | Sem_Diversity | 0.48 | FA1 | 6 | 0.34 | 9.19 | 2924 | 0.000 | 0.000 |
| Recog_Memory | PUP | 0.12 | FA2 | 9 | 0.02 | 8.68 | 2924 | 0.000 | 0.000 |
| Recog_Memory | Consistency_Token_FB_ON | 0.09 | FA3 | 1 | 0.10 | -0.94 | 2924 | 0.345 | 0.390 |
| Recog_Memory | Consistency_Token_FF_N | 0.02 | FA4 | 5 | 0.01 | 1.55 | 2924 | 0.121 | 0.158 |
| Recog_Memory | TrigramF_Avg_C_Log | 0.23 | FA5 | 6 | 0.07 | 7.83 | 2924 | 0.000 | 0.000 |
| Recog_Memory | Conc_Brys | 0.34 | FA6 | 2 | 0.25 | 9.83 | 2924 | 0.000 | 0.000 |
| Recog_Memory | Sem_N_D_Taxonomic_N3 | 0.37 | FA7 | 2 | 0.36 | 1.24 | 2924 | 0.216 | 0.261 |
| Recog_Memory | Interoceptive_Lanc | 0.16 | FA8 | 8 | 0.04 | 5.89 | 2924 | 0.000 | 0.000 |
| LexicalD_ACC_V_ELP | Sem_N_D | 0.36 | FA1 | 3 | 0.31 | 6.77 | 11626 | 0.000 | 0.000 |
| LexicalD_ACC_V_ELP | OUP | 0.12 | FA2 | 3 | 0.09 | 7.07 | 11626 | 0.000 | 0.000 |
| LexicalD_ACC_V_ELP | Consistency_Type_FB_N | 0.02 | FA3 | 9 | 0.00 | 2.24 | 11626 | 0.025 | 0.037 |
| LexicalD_ACC_V_ELP | Consistency_Token_FF_ON | 0.02 | FA4 | 3 | 0.02 | 1.57 | 11626 | 0.117 | 0.155 |
| LexicalD_ACC_V_ELP | TriphonP_St | 0.05 | FA5 | 12 | 0.00 | 7.76 | 11626 | 0.000 | 0.000 |
| LexicalD_ACC_V_ELP | AoA_Kuper | 0.19 | FA6 | 7 | 0.05 | 15.68 | 11626 | 0.000 | 0.000 |
| LexicalD_ACC_V_ELP | Sem_N_D_Taxonomic_N50 | 0.14 | FA7 | 1 | 0.16 | -5.30 | 11626 | 0.000 | 0.000 |
| LexicalD_ACC_V_ELP | Assoc_Freq_Type123 | 0.24 | FA8 | 5 | 0.17 | 8.16 | 11626 | 0.000 | 0.000 |
| LexicalD_ACC_V_ECP | Sem_N_D | 0.46 | FA1 | 3 | 0.40 | 6.20 | 7449 | 0.000 | 0.000 |
| LexicalD_ACC_V_ECP | OUP | 0.13 | FA2 | 3 | 0.10 | 3.71 | 7449 | 0.000 | 0.000 |
| LexicalD_ACC_V_ECP | Consistency_Type_FB_N | 0.02 | FA3 | 6 | 0.00 | 1.60 | 7449 | 0.110 | 0.148 |
| LexicalD_ACC_V_ECP | Consistency_Token_FF_ON | 0.02 | FA4 | 2 | 0.02 | 0.09 | 7449 | 0.929 | 0.936 |

| LexicalD_ACC_V_ECP | BigramF_Avg_U_Log | 0.05 | FA5 | 5 | 0.02 | 2.55 | 7449 | 0.011 | 0.017 |
| LexicalD_ACC_V_ECP | AoA_Kuper | 0.24 | FA6 | 5 | 0.10 | 12.29 | 7449 | 0.000 | 0.000 |
| LexicalD_ACC_V_ECP | Sem_N_D_Taxonomic_N50 | 0.27 | FA7 | 2 | 0.27 | 0.14 | 7449 | 0.886 | 0.909 |
| LexicalD_ACC_V_ECP | Assoc_Freq_Type123 | 0.35 | FA8 | 5 | 0.27 | 8.73 | 7449 | 0.000 | 0.000 |
| LexicalD_ACC_V_BLP | Sem_N_D | 0.40 | FA1 | 3 | 0.32 | 9.56 | 9608 | 0.000 | 0.000 |
| LexicalD_ACC_V_BLP | OUP | 0.22 | FA2 | 3 | 0.18 | 5.45 | 9608 | 0.000 | 0.000 |
| LexicalD_ACC_V_BLP | Consistency_Type_FB_N | 0.05 | FA3 | 1 | 0.05 | -1.20 | 9608 | 0.229 | 0.269 |
| LexicalD_ACC_V_BLP | Consistency_Type_FF_R | 0.05 | FA4 | 7 | 0.01 | 4.26 | 9608 | 0.000 | 0.000 |
| LexicalD_ACC_V_BLP | TriphonP_St | 0.08 | FA5 | 12 | 0.01 | 10.39 | 9608 | 0.000 | 0.000 |
| LexicalD_ACC_V_BLP | AoA_LWV | 0.20 | FA6 | 5 | 0.08 | 12.22 | 9608 | 0.000 | 0.000 |
| LexicalD_ACC_V_BLP | Sem_N_D_Taxonomic_N50 | 0.16 | FA7 | 1 | 0.18 | -4.73 | 9608 | 0.000 | 0.000 |
| LexicalD_ACC_V_BLP | Assoc_Freq_Type123 | 0.19 | FA8 | 4 | 0.16 | 2.92 | 9608 | 0.003 | 0.006 |
| LexicalD_ACC_A_MALD | Sem_N_D | 0.11 | FA1 | 4 | 0.06 | 5.24 | 10732 | 0.000 | 0.000 |
| LexicalD_ACC_A_MALD | OUP | 0.12 | FA2 | 1 | 0.12 | -0.80 | 10732 | 0.426 | 0.468 |
| LexicalD_ACC_A_MALD | Consistency_Type_FB_N | 0.03 | FA3 | 7 | 0.00 | 3.82 | 10732 | 0.000 | 0.000 |
| LexicalD_ACC_A_MALD | Consistency_Type_FF_R | 0.03 | FA4 | 6 | 0.00 | 2.69 | 10732 | 0.007 | 0.011 |
| LexicalD_ACC_A_MALD | TriphonP_Un | 0.05 | FA5 | 6 | 0.01 | 4.91 | 10732 | 0.000 | 0.000 |
| LexicalD_ACC_A_MALD | Interoceptive_Lanc | 0.05 | FA6 | 7 | 0.01 | 3.92 | 10732 | 0.000 | 0.000 |
| LexicalD_ACC_A_MALD | Sem_N_D_Taxonomic_N50 | 0.04 | FA7 | 1 | 0.05 | -0.98 | 10732 | 0.328 | 0.375 |
| LexicalD_ACC_A_MALD | Assoc_Freq_Type123 | 0.07 | FA8 | 3 | 0.07 | 0.67 | 10732 | 0.504 | 0.544 |
| LexicalD_ACC_A_AELP | Sem_N_D | 0.22 | FA1 | 8 | 0.13 | 8.71 | 4495 | 0.000 | 0.000 |
| LexicalD_ACC_A_AELP | PUP | 0.46 | FA2 | 1 | 0.47 | -1.74 | 4495 | 0.082 | 0.113 |
| LexicalD_ACC_A_AELP | Consistency_Token_FB | 0.06 | FA3 | 8 | 0.01 | 11.94 | 4495 | 0.000 | 0.000 |
| LexicalD_ACC_A_AELP | Consistency_Type_FF_R | 0.24 | FA4 | 9 | 0.00 | 15.81 | 4495 | 0.000 | 0.000 |
| LexicalD_ACC_A_AELP | TriphonP_St | 0.26 | FA5 | 5 | 0.19 | 6.92 | 4495 | 0.000 | 0.000 |
| LexicalD_ACC_A_AELP | Conc_Brys | 0.21 | FA6 | 3 | 0.16 | 6.63 | 4495 | 0.000 | 0.000 |
| LexicalD_ACC_A_AELP | Sem_N_D_Taxonomic_N50 | 0.22 | FA7 | 5 | 0.22 | 1.32 | 4495 | 0.188 | 0.233 |
| LexicalD_ACC_A_AELP | Assoc_Freq_Type123 | 0.21 | FA8 | 6 | 0.17 | 3.59 | 4495 | 0.000 | 0.001 |
| Naming_ACC_ELP | AoA_LWV | 0.21 | FA1 | 1 | 0.22 | -1.19 | 11617 | 0.235 | 0.274 |
| Naming_ACC_ELP | NSyll | 0.13 | FA2 | 9 | 0.10 | 7.46 | 11617 | 0.000 | 0.000 |
| Naming_ACC_ELP | Consistency_Token_FB_ON | 0.04 | FA3 | 5 | 0.02 | 2.16 | 11617 | 0.030 | 0.045 |
| Naming_ACC_ELP | Consistency_Token_FF_R | 0.12 | FA4 | 5 | 0.10 | 2.44 | 11617 | 0.015 | 0.022 |
| Naming_ACC_ELP | BiphonP_Un | 0.06 | FA5 | 1 | 0.07 | -1.64 | 11617 | 0.101 | 0.138 |
| Naming_ACC_ELP | AoA_LWV | 0.21 | FA6 | 4 | 0.10 | 12.33 | 11617 | 0.000 | 0.000 |
| Naming_ACC_ELP | Sem_N_D_Taxonomic_N50 | 0.05 | FA7 | 1 | 0.06 | -1.21 | 11617 | 0.225 | 0.269 |
| Naming_ACC_ELP | Assoc_Freq_Type123 | 0.16 | FA8 | 5 | 0.10 | 6.11 | 11617 | 0.000 | 0.000 |
| SemanticD_ACC_Calgary | Nsenses_WordNet | 0.15 | FA1 | 15 | 0.03 | 6.28 | 2427 | 0.000 | 0.000 |
| SemanticD_ACC_Calgary | BigramF_Avg_C_Log | 0.19 | FA2 | 3 | 0.18 | 0.52 | 2427 | 0.600 | 0.629 |
| SemanticD_ACC_Calgary | Consistency_Token_FB_N | 0.13 | FA3 | 1 | 0.14 | -1.13 | 2427 | 0.258 | 0.298 |
| SemanticD_ACC_Calgary | Consistency_Type_FF_R | 0.11 | FA4 | 7 | 0.05 | 2.73 | 2427 | 0.006 | 0.010 |
| SemanticD_ACC_Calgary | TrigramF_Avg_C_Log | 0.16 | FA5 | 6 | 0.06 | 4.71 | 2427 | 0.000 | 0.000 |
| SemanticD_ACC_Calgary | Mink_Perceptual_Lanc | 0.15 | FA6 | 3 | 0.14 | 0.70 | 2427 | 0.483 | 0.527 |
| SemanticD_ACC_Calgary | Sem_N_D_Taxonomic_N50 | 0.04 | FA7 | 3 | 0.03 | 1.43 | 2427 | 0.154 | 0.199 |
| SemanticD_ACC_Calgary | Assoc_Freq_Token123 | 0.16 | FA8 | 4 | 0.12 | 2.13 | 2427 | 0.033 | 0.048 |

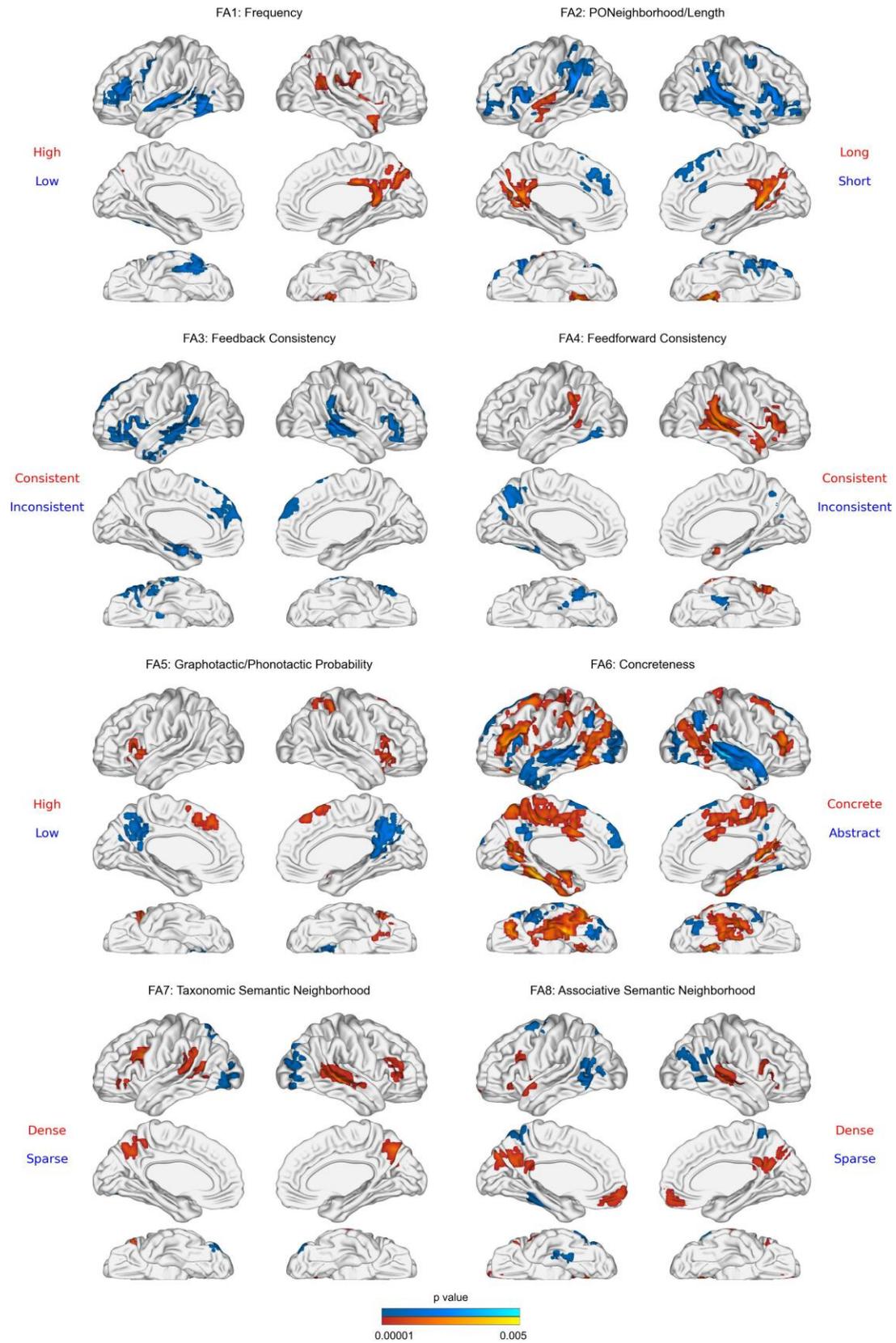

***Supplementary Fig. 1 | Voxel-wise modulation maps showing how BOLD was
modulated by each latent factor after controlling for the other factors.***

*Notes*. $p < .005$ at voxel-wise, cluster-size based correction, $p < .05$ at whole-brain level.

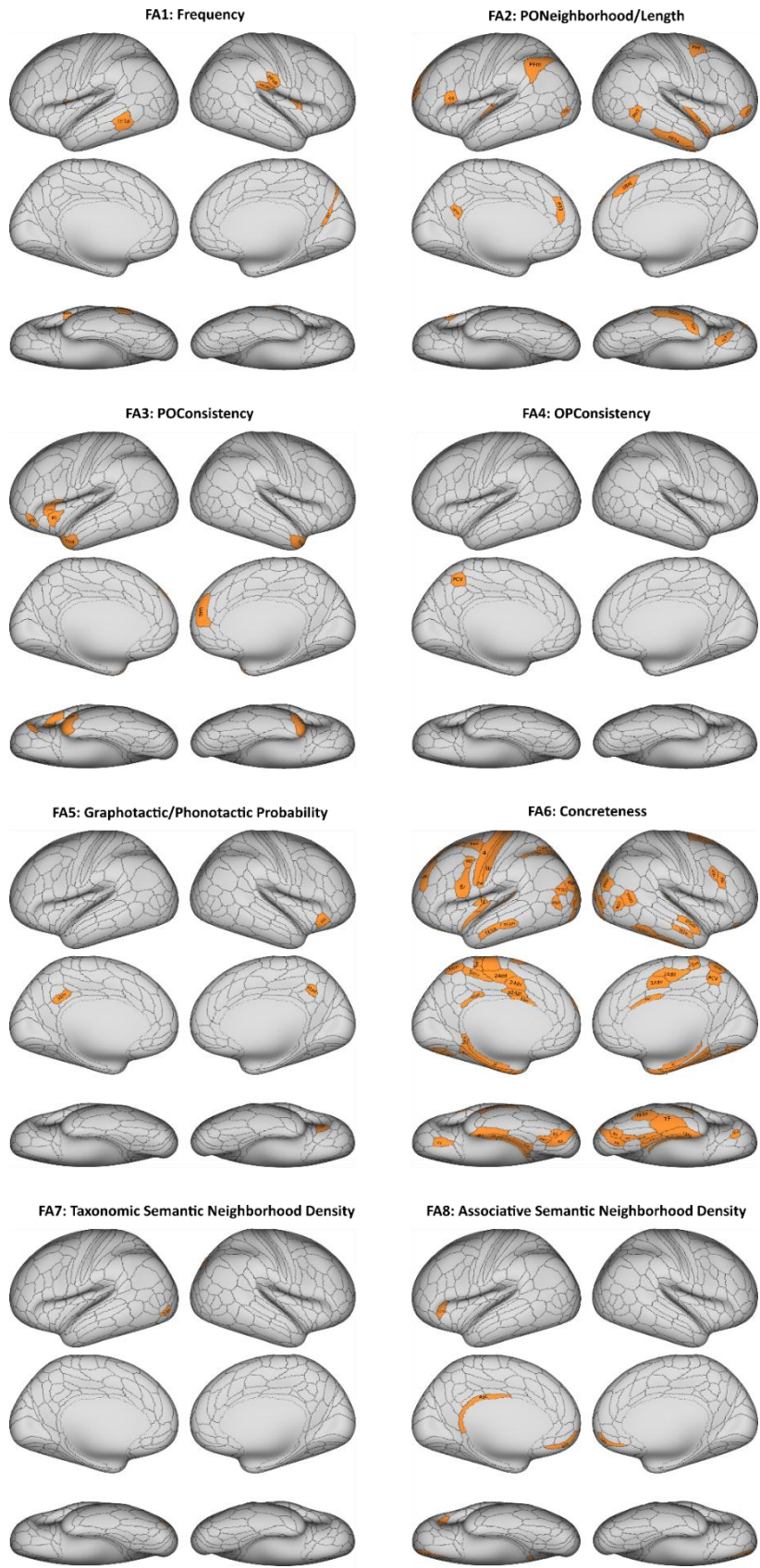

**Supplementary Fig. 2 | Parcel-wise modulation maps showing the parcels only significantly modulated by one factor but not the others.**

*Notes.* $p < .005$ at parcel-wise, FDR corrected.

**Supplementary references**